\documentclass[10pt]{article}

\usepackage{amsmath}
\usepackage{amsfonts}
\usepackage{amssymb}
\usepackage{mathrsfs}
\usepackage{latexsym, amscd, amsthm}
\usepackage[english]{babel}

\DeclareFixedFont{\sfracFont}{U}{euf}{b}{n}{7pt}
\newtheoremstyle{mydefi}
  {15pt}
  {15pt}
  {}
  {}
  {\bfseries}
  {:}
  {.5em}
  {}
\newtheoremstyle{mytheo}
  {15pt}
  {15pt}
  {\slshape}
  {}
  {\bfseries}
  {:}
  {.5em}
  {}

\theoremstyle{mytheo}
\newtheorem{stel}{Theorem}[section]
\newtheorem{lem}[stel]{Lemma}

\theoremstyle{mydefi}
\newtheorem{de}[stel]{Definition}

\newcommand{\BB}[1]{\mathbb{#1}}
\newcommand{\BS}[1]{\mathscr{#1}}

\newcommand{\p}{|\!|}
\newcommand{\ten}{\otimes}
\newcommand{\AC}{\mathcal{A}}
\newcommand{\BC}{\mathcal{B}}
\newcommand{\CC}{\mathcal{C}}
\newcommand{\DC}{\mathcal{D}}

\newcommand{\FC}{\mathcal{F}}
\newcommand{\HC}{\mathcal{H}}

\newcommand{\LC}{\mathcal{L}}

\newcommand{\NC}{\mathcal{N}}

\newcommand{\WC}{\mathcal{W}}

\begin{document}
\title{Quantum filtering: a reference probability approach}
\author{Luc Bouten and Ramon van Handel \\ \\
\normalsize{Physical Measurement and Control 266-33, California Institute 
of Technology,}\\ \normalsize{1200 E.\ California Blvd., Pasadena, CA 
91125, USA}} \date{}
\maketitle

\vspace{8mm}

\begin{abstract} 
These notes are intended as an introduction to noncommutative (quantum)
filtering theory. An introduction to quantum probability theory is given,
focusing on the spectral theorem and the conditional expectation as the
least squares estimate, and culminating in the construction of Wiener and
Poisson processes on the Fock space.  Next we describe the
Hudson-Parthasarathy quantum It\^o calculus and its use in the modelling of
physical systems.  Finally, we use a reference probability method to
obtain quantum filtering equations, in the Belavkin-Zakai (unnormalized)
form, for several system-observation models from quantum optics.  The
normalized (Belavkin-Kushner-Stratonovich) form is obtained through a
noncommutative analogue of the Kallianpur-Striebel formula. 
\end{abstract}

\section{Introduction}\label{sec introduction}

The development of a theory of feedback control in the 1950s and 60s has
provided a huge stimulus to the engineering disciplines.  Nowadays all but
the simplest of devices that make up our everyday lives use feedback to
provide efficient and reliable performance despite the ever increasing
complexity and miniaturization.  However, at a time when microtechnology
is making way for nanotechnology, we are rapidly approaching the boundary
of the classical world past which the effects of quantum mechanics cannot
be neglected.  The theory of quantum mechanics tells us that any
description of the phenomena at small scales is inherently
nondeterministic in nature. This opens new areas of application for
stochastic control theory, which will likely play an important role in a
future generation of technology.  In particular, as observations of
quantum systems are inherently noisy, the theory of filtering---the
extraction of information from a noisy signal---forms an essential part of
any quantum feedback control strategy. 

These notes are intended as an introduction to quantum filtering theory.
We begin by giving a brief overview of noncommutative probability theory
\cite{Maa03,Bia95}, the associated stochastic calculus \cite{HuP84}, and
the role of this theory in the modelling of quantum dynamical systems
\cite{AFLu90, Gou04, Gou05}.  We then obtain quantum filtering equations
\cite{Bel92b} for a variety of quantum models.  We systematically use
change of measure techniques, bringing our approach close to the classical
reference probability method of Zakai \cite{Zak69}. 

The study of quantum filtering and control was pioneered by V.P.~Belavkin
in a remarkable series of articles, dating back to the early 1980s
\cite{Bel80,Bel83,Bel87,Bel92a,Bel92b,Bel97}. A discrete version of his
filtering equation can already be found in \cite{Bel80}.  After the
development of quantum stochastic calculus \cite{HuP84}, continuous time
versions of Belavkin's equation appeared in \cite{Bel87}, \cite{Bel92a}
and \cite{Bel92b}.  The sample paths of the solution of the quantum
filtering equation \cite{Dav69,Dav76,Car93,WiM93a} are known in the
continuous time quantum measurement community as quantum trajectories. 

The reference probability method used here is inspired by the classical
treatment of Zakai \cite{Zak69}.  In this approach we change the
underlying measure so that the filtering problem reduces to elementary
manipulations of the conditional expectation with respect to the new
(reference) measure. The key idea that allows us to apply this approach to
the quantum case is the observation that the Radon-Nikodym derivative must
be compatible with the observations that we are conditioning on.  To
obtain a change of measure that satisfies this property we employ a
technique that to our knowledge first appeared in a paper by Holevo
\cite{Hol90} (see also \cite{Bel92a,GG94,Bar96}), which replaces here the
Girsanov tranformation in Zakai's treatment.  Apart from this point, the
derivation is essentially classical, i.e.\ we derive the Belavkin-Zakai
equation and obtain the (nonlinear) Belavkin-Kushner-Stratonovich equation
via a noncommutative analogue of the Kallianpur-Striebel formula.

In contrast to the martingale techniques used in \cite{Bel92b, BGM04}, the
reference probability method is independent from the innovations
conjecture \cite{DM81} (as is the case in classical filtering).  Moreover,
the reference probability approach consists only of elementary
manipulations of the (quantum) conditional expectation. As in classical
filtering theory the Kallianpur-Striebel formula has found a wide range of
applications, it is our hope that a similar approach will be fruitful in
the quantum context and will clarify some of the existing literature on
quantum filtering theory. 

After the filtering equations have been obtained, methods from classical
nonlinear stochastic control can be applied \cite{Bel83,HSM05,BEB05} to
design control laws.  Thus the presence of quantum mechanics in quantum
feedback control remains limited to the filtering procedure. Recent
experiments implementing quantum feedback controls
\cite{AASDM02,GSDM03,GSM04} have led to renewed interest in the field
which is now rapidly expanding
\cite{DHJMT00,HSM05,BEB05,Jam04,Jam05,GBS05, EdB05,HaM05,HSM05a,BGM04}. 
The interaction between the areas of stochastic control and theoretical
and experimental physics is essential in paving the way towards the
engineering of quantum technologies. 

There are three key ingredients that are required for the development of
quantum filtering theory.  First, we need to capture both classical
probability and quantum mechanics within the framework of a generalised
probability theory, called noncommutative or quantum probability theory. 
The central object in this theory, the spectral theorem, allows us to make a 
seamless connection between quantum systems and the associated classical 
probabilistic measurement outcomes.  Second, we need a noncommutative 
generalization of the concept of conditional expectations.  As in 
classical probability, we will find that a suitably restricted definition 
of the quantum conditional expectation is none other than a least squares 
estimator, which elucidates its role in quantum filtering theory.  Finally,
we need a noncommutative analogue of a stochastic calculus and of 
stochastic differential equations.  This construction provides a 
broad class of models for which we can obtain quantum filtering equations.

In the following we develop each of these themes in turn.  Section 
\ref{sec QP} introduces noncommutative probability theory, focusing on 
the spectral theorem, and introduces the quantum conditional expectation as 
a least squares estimate.  In section \ref{sec Fock} we show how Wiener 
and Poisson processes emerge in a particular quantum probabilistic model
based on the Fock space.  The construction additionally demonstrates the 
use of the spectral theorem.  Section \ref{sec qsc} is an exposition of the
noncommutative stochastic calculus of Hudson and Parthasarathy.
In Section \ref{sec filtering} we introduce the system-observation models 
which we are interested in; they are obtained from physical models via a 
weak coupling limit. Section \ref{sec KS formula} deals with the 
derivation of the Belavkin-Zakai and the Belavkin-Kushner-Stratonovich 
equation using the reference probability approach.  We conclude these notes 
with some additional examples.

\section{Noncommutative probability theory}\label{sec QP}

In this section we wish to make the point that quantum mechanics is a
probability theory in which the random variables, called observables in
quantum mechanics, are allowed to be noncommutative. Indeed, in quantum
theory observables are represented by self-adjoint operators on some
Hilbert space $\BS{H}$ which in general need not commute. In order to
incorporate quantum mechanics in the framework of probability theory we
will weaken the axioms of the latter to allow for noncommutative random
variables.  By considering commuting observables only, classical
probability theory can be recovered using the spectral theorem.  The
latter asserts that commuting self-adjoint operators can be diagonalised
simultaneously, allowing them to be interpreted simultaneously as
functions (i.e.\ random variables) on the diagonal (the spectrum).
Excellent introductions to noncommutative probability theory are
\cite{Maa03,Bia95}, see also \cite{Bou04b,HSM05}. 

As quantum probability is algebraic in nature, it is instructive to begin 
by forming an algebraic picture of classical probability theory.  
Consider a classical probability space $(\Omega,\Sigma,{\bf P})$.  Then 
the space $\mathcal{A}=L^\infty(\Omega,\Sigma,{\bf P})$ of bounded 
measurable complex functions on $\Omega$ (we will always use {\it complex} 
function spaces $L^\infty$, $L^2$ etc.) is a $^*$-algebra: any complex 
linear combination, pointwise multiplication, and conjugate of 
functions in $\mathcal{A}$ are still in $\mathcal{A}$.  We can use 
integration with respect to ${\bf P}$ to map any element of 
$\mathcal{A}$ to a complex number.  If we operate this map 
$\mathbb{P}:\mathcal{A}\to\mathbb{C}$ on an indicator function, we obtain 
the probability of the corresponding event.  As the algebra only contains 
measurable functions, and as it is in fact generated by the 
indicator functions (any element in $L^\infty$ can be approximated 
arbitrarily well by linear combinations of its indicator functions), the 
pair $(\mathcal{A},\mathbb{P})$ encodes the same information as the 
probability space $(\Omega,\Sigma,{\bf P})$ (at least up to sets of 
measure zero \cite{Maa03}, a point on which we will not dwell further.)
It is convenient to represent the elements of the algebra
$\mathcal{A}=L^\infty(\Omega,\Sigma,{\bf P})$ as {\it operators} on the
Hilbert space $\BS{H}=L^2(\Omega,\Sigma,{\bf P})$.  This can be done by
letting $\mathcal{A}$ act on $\BS{H}$ via pointwise multiplication (note
that $A\psi$ is an element of $\BS{H}$ for any $A\in\mathcal{A}$,
$\psi\in\BS{H}$).  

Let us now generalize these ideas.  Denote by $\BC(\BS{H})$ the algebra of
all bounded operators on a Hilbert space $\BS{H}$, and let
$\mathcal{S}\subset\BC(\BS{H})$. The set $\mathcal{S}':= \{X \in
\BC(\BS{H});\ XS = SX,\ \forall S\in\mathcal{S}\}$ is called the
\emph{commutant} of $\mathcal{S}$ in $\BC(\BS{H})$. A \emph{von Neumann
algebra} $\AC$ on $\BS{H}$ is a $^*$-subalgebra of $\BC(\BS{H})$ such that
$\AC$ equals its double commutant, i.e.\ $\AC''=\AC$. Von Neumann's double
commutant theorem then asserts that $\AC$ is closed in the weak operator
topology, i.e.\ if for some net $\{A_j\}\in \AC$ the inner products
$\langle x,A_jy\rangle \to \langle x, Ay\rangle$ for all $x,y \in \BS{H}$,
then $A$ is an element of $\AC$.  This property guarantees that
a von Neumann algebra is generated by its projections $P\in\AC$,
$P=P^2=P^*$ \cite{KaR83}. 

From $\AC'' =\AC$ it immediately follows that the identity $I$ is an
element of the von Neumann algebra $\AC$. A \emph{state} on $\AC$ is a
linear map $\mathbb{P}:\ \AC \to \BB{C}$ such that $\mathbb{P}$ is 
\emph{positive} in the sense that $\mathbb{P}(A^*A) \ge 0$ for all $A \in 
\AC$ and such that $\mathbb{P}$ is normalised, i.e.\ $\mathbb{P}(I) = 1$. A 
state is called \emph{normal} if it is weak operator continuous on the 
unit ball of $\AC$. It is called \emph{faithful} if $\mathbb{P}(A^*A)=0$ 
implies $A = 0$.  We remark that $L^\infty(\Omega,\Sigma,{\bf P})$ is a 
von Neumann algebra on $L^2(\Omega,\Sigma,{\bf P})$ with normal state 
$\mathbb{P}$.

The following theorem (see \cite{KaR86} sections 9.3--9.5 for a proof) is at 
the heart of spectral theory. 

\begin{stel}\label{thm spectral}\textbf{(Spectral Theorem)}
Let $\CC$ be a commutative von Neumann algebra and $\mathbb{P}$ a normal
state on $\CC$. Then there is a probability space $(\Omega, \Sigma, {\bf P})$
such that $\CC$ is $^*$-isomorphic to $L^\infty(\Omega, \Sigma, {\bf P})$,
the space of all bounded measurable functions on $\Omega$. Furthermore, if
we denote the $^*$-isomorphism by $\iota:\ \CC \to L^\infty(\Omega,
\Sigma, {\bf P})$ then we have
  \begin{equation*}
  \mathbb{P}(C) = \int_\Omega \iota(C){\bf P}(d\omega),\ \ \ \ C \in \CC.
  \end{equation*}
\end{stel}  

{\bf Example.} We are guided by the elementary case of a Hilbert space
with dimension $n<\infty$.  Any linear operator on such a space can be
represented as a (complex) $n\times n$ matrix, and any $^*$-algebra of
matrices is a von Neumann algebra.  Now consider a set of matrices
$\{A_i\}$ that commute with each other and with all their adjoints,
$[A_i,A_j]=[A_i,A_j^*]=0$ $\forall i,j$.  These matrices generate a
commutative von Neumann (matrix) algebra $\mathcal{C}$.  We can now
simultaneously diagonalize every element of $\mathcal{C}$ using a unitary
$U$.  To each $A\in\mathcal{C}$, we associate a map $\iota(A): \{1,\ldots,
n\}\to\mathbb{C}$ such that $\iota(A)(i)$ is the $i$th diagonal element of
the diagonalized matrix $U^*AU$.  Then $\iota(A)\in L^\infty(\{1,\ldots,
n\})$, and $\iota(AB)=\iota(A)\iota(B)$ under pointwise multiplication in
$L^\infty$.  Now choose $\Sigma=\sigma(\iota(C);\ C\in\mathcal{C})$, the
sigma-algebra generated by $\iota(\mathcal{C})$.  Finally, given a state
$\mathbb{P}$ on $\mathcal{C}$, define for $\Gamma\in\Sigma$ the measure
${\bf P}(\Gamma)=\mathbb{P}(UC^\Gamma U^*)$ where
$C^\Gamma_{ii}=\chi_\Gamma(i)$ ($\chi_\Gamma$ is the indicator function
of $\Gamma$).  Then we have explicitly constructed a $^*$-isomorphism
$\iota:\mathcal{C}\to L^\infty(\{1,\ldots,n\},\Sigma,{\bf P})$.  The
spectral theorem is an extension of these ideas to infinite dimensional
operators. Though technically much more involved, the flavour of the
procedure remains the same. 

We have already described how a classical probability space can be
encoded algebraically by a commutative von Neumann algebra with normal
state.  The spectral theorem allows us to conclude that studying
commutative von Neumann algebras equipped with normal states is 
equivalent to studying probability spaces. This motivates the definition of a
\emph{noncommutative} or \emph{quantum} probability space as a von Neumann
algebra $\NC$ equipped with a normal state $\mathbb{P}$.  The events in this
theory are the projections in $\NC$, and the state plays the role of a
probability measure.  We can now see how the various technical properties
of von Neumann algebras contribute to their interpretation as
probabilistic models. Weak closure guarantees that the theory is completely
determined by the set of all events; and normality of the state is
equivalent to countable additivity \cite{KaR86}. 

Any physical experiment can be described by classical probability,
provided that we consistently perform the same measurements.  For example,
if we measure the position of a particle at time $t$, and repeat this
experiment many times with the same initial conditions, then the
statistics of the measurement outcomes are entirely described by a
classical probabilistic model.  The break with classical probability
occurs because in quantum mechanics there exist observables, such as
position and momentum, that cannot be measured simultaneously in the same
experimental realization.  Such measurements are said to be {\it
incompatible}, and this is enforced by the fact that they do not commute. 
The spectral theorem provides a concrete mathematical implementation of
these ideas: a set of compatible events generates a commutative algebra,
and is hence equivalent to a classical probabilistic model.  Thus if $E$
and $F$ are compatible events we can interpret $EF$ as the event $\langle
E$ and $F\rangle$ and $E\vee F := E +F - EF$ as the event $\langle E$ or
$F\rangle$. However, two incompatible events are not represented on the
same classical probability space, and their joint probability is
undefined. 

Having considered the noncommutative analogue of a probability space, let
us now turn our attention to random variables.  Let us first investigate a
classical random variable $f:\Omega\to\mathbb{R}$.  If $f$ is bounded, we
could interpret it as an element of the algebra
$\mathcal{N}=L^\infty(\Omega,\Sigma,{\bf P})$.  In general, however, a
random variable need not be bounded.  Nonetheless we can still consider $f$
as a (self-adjoint) operator acting on $\BS{H}=L^2(\Omega,\Sigma,{\bf P})$
by pointwise multiplication, provided that we restrict its domain to the
set of $\psi\in\BS{H}$ such that $f\psi\in\BS{H}$.  To make the algebraic
description self-contained, we must now express the fact that $f$ is
measurable in terms of the algebra $\mathcal{N}$.  This can be done by
requiring that for every Borel set $B\subset\mathbb{R}$ the indicator
function $\chi_{\{\omega\in\Omega:f(\omega)\in B\}}$, considered as an
operator on $\BS{H}$ by pointwise multiplication, is an element of
$\mathcal{N}$. 

A random variable on a quantum probability space, called an {\it 
observable}, is a self-adjoint operator on $\BS{H}$.  If an observable 
$F$ is an element of $\mathcal{N}$ (and hence is bounded), then it 
generates a commutative von Neumann algebra and we can use the spectral 
theorem to map it to a random variable $\iota(F)$ on a classical 
probability space.  In general, however, $F$ need not be bounded; we must 
extend the map $\iota$ to be able to represent unbounded observables as 
classical random variables.  To this end, note that since $F$ is 
self-adjoint its spectrum is real.  Therefore we can define two bounded 
commuting operators $T_+ := (F+iI)^{-1}$ and $T_-:=(F-iI)^{-1}$. The 
operators $T_+$ and $T_-$ generate a commutative von Neumann algebra, and 
can hence be represented on a classical probability space by the spectral 
theorem. We now extend $\iota$ as
  \begin{equation*}
  \iota(F) := \frac{\big(\iota(T_+)\big)^{-1} + \big(\iota(T_-)\big)^{-1}}{2}.  
  \end{equation*}
Then $\iota(F)$ is the representation of $F$ as a classical random 
variable.  Now define the {\it spectral measure}
$$
	E(B)=\iota^{-1}(\chi_{\{\omega\in\Omega:\iota(F)(\omega)\in B\}})
$$
for any Borel set $B$ of $\mathbb{R}$.  $E(B)$ is the event in 
$\mathcal{N}$ corresponding to $\langle F$ takes a value in $B\rangle$.  
$F$ is {\it affiliated} to a von Neumann algebra $\mathcal{C}$ if for all 
Borel sets $B$ we have $E(B)\in\mathcal{C}$.  This concept is equivalent 
to measurability of a classical random variable.  Finally, note that if the 
observable $F$ is an element of the algebra, then the expectation of $F$ 
is given by $\mathbb{P}(F)$.  This is consistent with the interepretation 
of $F$ as a classical random variable $\iota(F)$, as 
$\mathbb{P}(F)=\mathbb{E}_{\bf P}[\iota(F)]$.  Using the extension of 
$\iota$ we can extend the state $\mathbb{P}$ to unbounded observables by
$$
	\mathbb{P}(F)=\mathbb{E}_{\bf P}[\iota(F)]=
	\int_\mathbb{R}\lambda\,\mathbb{P}(E(d\lambda))
$$
Then $\mathbb{P}(F)$ is the expectation of the unbounded observable $F$.

\begin{de}\label{de conditional expectation}\textbf{(Conditional expectation)}
Let $(\mathcal{A},\mathbb{P})$ be a quantum probability space and let
$\mathcal{C}\subset\mathcal{Z}=\{Z\in\mathcal{A};\ AZ=ZA,\ 
\forall\,A\in\mathcal{A}\}$, the {\it center} of $\mathcal{A}$.  Then
$\mathbb{P}(\cdot|\mathcal{C}):\mathcal{A}\to\mathcal{C}$ is 
(a version of) the {\it conditional expectation} from $\mathcal{A}$ onto 
$\mathcal{C}$ if for all $A\in\mathcal{A}$, we have
$\mathbb{P}(\mathbb{P}(A|\mathcal{C})C)=\mathbb{P}(AC)$ $\forall 
C\in\mathcal{C}$.
\end{de}

The center $\mathcal{Z}$ is by definition a commutative algebra.  In
applications, we begin with a fixed probability space
$(\mathcal{N},\mathbb{P})$ and specify a commutative von Neumann
subalgebra $\mathcal{Z}\subset\mathcal{N}$ generated by the observations
we have performed.  We then choose $\mathcal{A}=\mathcal{Z}'$, the
commutant of $\mathcal{Z}$ in $\mathcal{N}$.  By the double commutant
theorem, $\mathcal{Z}$ is the center of $\mathcal{Z}'$.  Def.\ \ref{de
conditional expectation} only allows conditioning of operators that
commute with all the elements of the commutative algebra $\mathcal{Z}$. 
This is a natural requirement, as there is no need for updating
observables that are incompatible with what has already been observed.
Note that $\iota(\mathbb{P}(A|\mathcal{C}))=\mathbb{E}_{\bf
P}[\iota(A)|\sigma(\iota(C);\ C\in\mathcal{C})]$ for any self-adjoint
$A\in\mathcal{A}$, where $\iota$ is given by the spectral theorem applied
to the commutative von Neumann algebra generated by $A$ and $\mathcal{C}$.
Hence for observables the conditional expectation coincides with its
classical counterpart. 

{\bf Remark.} We have taken a more restrictive definition of the
conditional expectation than is usual in quantum probability, i.e.\ we do
not allow for conditioning on noncommutative algebras.  The more general
definition \cite{Tak71} does not have a direct physical interpretation but
is required for the definition of concepts such as a noncommutative Markov
process \cite{Kum85}.  Beside its direct physical meaning, one advantage
of our more restrictive approach is that a conditional expectation from
$\AC$ onto $\CC$ in the above sense always exists (see \cite{BGM04} for a
``construction'' using the central decomposition of $\AC$) and is unique
with probability one.  In fact, our conditional expectation is a special
case of the conditional expectation of \cite{Tak71}. 

\begin{lem}
	The conditional expectation
	of Def.\ \ref{de conditional expectation} exists and is unique with 
	probability one {\rm(}i.e., any two versions $P$ and $Q$ of 
	$\mathbb{P}(A|\mathcal{C})$ satisfy $\|P-Q\|_\mathbb{P}=0$, where
	$\|X\|^2_\mathbb{P}:=\mathbb{P}(X^*X)$.{\rm)}
	Moreover, $\mathbb{P}(A|\mathcal{C})$ is the {\rm least mean 
	square} estimate of $A$ given $\mathcal{C}$, i.e.\
	$\|A-\mathbb{P}(A|\mathcal{C})\|_\mathbb{P}\le \|A-C\|_\mathbb{P}$ 
	$\forall\,C\in\mathcal{C}$.
\end{lem}

\begin{proof}
\begin{enumerate}
\item	{\it Existence.}
	We have already established that for self-adjoint $A\in\mathcal{A}$,
	we can explicitly define a $\mathbb{P}(A|\mathcal{C})$ that 
	satisfies the conditions of  Def.\ \ref{de conditional expectation}
	using the spectral theorem, i.e.\ $\mathbb{P}(A|\mathcal{C})=\iota^{-1}(\mathbb{E}_{\bf
	P}[\iota(A)|\sigma(\iota(C);\ C\in\mathcal{C})])$.  The classical 
	conditional expectation exists, and moreover the conditional 
	expectation of a bounded random variable is bounded.  Hence 
	$\mathbb{P}(A|\mathcal{C})$ exists in $\mathcal{C}$ for self-adjoint 
	$A\in\mathcal{A}$.  But any $A\in\mathcal{A}$ can be written as 
	$A=A_1+iA_2$ with self-adjoint $A_1=(A+A^*)/2$ and $A_2=i(A^*-A)/2$. As 
	$\mathbb{P}(A_1|\mathcal{C})$ and $\mathbb{P}(A_2|\mathcal{C})$
	exist and $\mathbb{P}(A|\mathcal{C})=\mathbb{P}(A_1|\mathcal{C})
	+i\mathbb{P}(A_2|\mathcal{C})$ satisfies the conditions of Def.\ 
	\ref{de conditional expectation}, existence is proved.
\item   {\it Uniqueness w.p.\ one.}   Define the pre-inner product $\langle X, 
	Y\rangle := \mathbb{P}(X^*Y)$ on $\AC$ (it might have nontrivial kernel if 
	$\mathbb{P}$ is not faithful.)  Then $\langle C,A-\BB{P}[A|\CC]\rangle =
	\mathbb{P}(C^*A)-\mathbb{P}(C^*\BB{P}[A|\CC])=0$
	for all $C \in \CC$ and $A \in \AC$, i.e.\ $A -\BB{P}[A|\CC]$ is 
	orthogonal to $\CC$.  Now let $P$ and $Q$ be two versions of
	$\BB{P}[A|\CC]$.  It follows that $\langle C,P-Q\rangle=0$
	for all $C\in\CC$.  But $P-Q\in\CC$, so $\langle P-Q,P-Q\rangle=
	\|P-Q\|_\mathbb{P}^2=0$.
\item	{\it Least squares.}  Note that for all $K\in\mathcal{C}$
  \begin{equation*}
  \|A-K\|_\mathbb{P}^2  = 
  \big\|A-\BB{P}[A|\CC]+ \BB{P}[A|\CC]-K\big\|_\mathbb{P}^2 
   = \big\|A-\BB{P}[A|\CC]\big\|_\mathbb{P}^2 + 
	\big\|\BB{P}[A|\CC]-K\big\|_\mathbb{P}^2 \ge
  \big\|A-\BB{P}[A|\CC]\big\|_\mathbb{P}^2,
  \end{equation*}
	where, in the next to last step, we used that $A-\BB{P}[A|\CC]$ is 
	orthogonal to $\BB{P}[A|\CC]-K\in \CC$. 
\end{enumerate}
\end{proof}

{\bf Remark.} The usual elementary properties of classical conditional
expectations and their proofs \cite{Wil91} carry over directly to the
noncommutative situation.  In particular, we have linearity, positivity,
invariance of the state
$\mathbb{P}(\mathbb{P}(A|\mathcal{C}))=\mathbb{P}(A)$, invariance of
$\mathcal{C}$ ($\mathbb{P}(A|\mathcal{C})=A$ if $A\in\mathcal{C}$), the
tower property $\mathbb{P}(\mathbb{P}(A|\mathcal{B})|\mathcal{C})=
\mathbb{P}(A|\mathcal{C})$ if $\mathcal{C}\subset\mathcal{B}$, the module
property $\mathbb{P}(AB|\mathcal{C})=B\mathbb{P}(A|\mathcal{C})$ for
$B\in\mathcal{C}$, etc.  As an example, let us prove linearity.  It
suffices to show that
$Z=\alpha\mathbb{P}(A|\mathcal{C})+\beta\mathbb{P}(B|\mathcal{C})$
satisfies the definition of $\mathbb{P}(\alpha A+\beta B|\mathcal{C})$,
i.e.\ $\mathbb{P}(ZC)=\mathbb{P}((\alpha A+\beta B)C)$ for all
$C\in\mathcal{C}$.  But this is immediate from the linearity of $\mathbb{P}$
and Definition \ref{de conditional expectation}. 

We reemphasize that the conditional expectation $\BB{P}[A|\CC]$ is a
\emph{least squares estimate} of $A$.  Hence the quantum filtering problem
is essentially an estimation problem, just like its classical counterpart. 
We can extend the conditional expectation, as we did for the state
$\mathbb{P}$, to be defined for any self-adjoint operator that is
affiliated to $\mathcal{A}$.  This way the least squares estimate given a
set of observations is defined for any quantum observable that is
compatible with these observations.

\section{Stochastic processes on Fock space}\label{sec Fock}

After having briefly discussed the framework of noncommutative probability
theory, we now turn to one particular quantum probability space.  Within
this model we will discover many interesting classical stochastic
processes, i.e.\ a whole family of Wiener processes and Poisson processes.
However, these processes do not commute amongst each other. An extension
of It\^o's stochastic calculus, due to Hudson and Parthasarathy
\cite{HuP84}, unites all these processes again in one noncommutative
stochastic calculus (see Section \ref{sec qsc}).  In Section \ref{sec
filtering} we shall argue that the model studied here appropriately
describes the quantised electromagnetic field and its interaction with
matter. 

Let $\BS{H}$ be a Hilbert space. The \emph{symmetric} or 
\emph{Bosonic Fock space over $\BS{H}$} is defined as
  \begin{equation*}
  \FC(\BS{H}) := \BB{C} \oplus \bigoplus_{n=1}^\infty \BS{H}^{\ten_s n}.
  \end{equation*}
We will take $\BS{H}$ to be $L^2(\BB{R})$, the space of quadratically 
integrable functions on $\BB{R}$, and denote $\FC\big(L^2(\BB{R})\big)$ 
simply by $\FC$. The Fock space $\FC$ is closely related to the Wiener 
chaos expansion in probability theory \cite{Mey93,Bia95} and from a 
physics point of view it describes a field of bosonic particles, like 
photons. Then the term $L^2(\BB{R})^{\ten_s n}$ in the direct sum 
defining $\FC$, describes the situation where there are $n$ photons 
present. Since photons are bosons they have to be described by symmetric 
wavefunctions, explaining the symmetric tensor product in the definition. 

For every $f \in \BS{H}$ we define the \emph{exponential vector} 
$e(f)\in \FC$ by
  \begin{equation}\label{eq exp vector}
  e(f) := 1 \oplus \bigoplus_{n=1}^\infty \frac{1}{\sqrt{n!}} f^{\ten n}.
  \end{equation}
The linear span $\DC$ of all exponential vectors is a dense 
subspace of $\FC$. On the dense domain $\DC$ we define for all 
$f\in \BS{H}$ an operator $W(f)$ by
  \begin{equation}\label{eq Weyl}
  W(f)e(g) := \exp\big(-\langle f, g\rangle - \frac{1}{2} \p f\p^2\big) e(f+g),\ \ \ \ g \in \BS{H}.
  \end{equation}
These operators are isometric and therefore uniquely extend 
to unitary operators, also denoted $W(f)$, on $\FC$. The 
operators $W(f):\ \FC \to \FC$ are called \emph{Weyl operators} 
and they satisfy the following \emph{Weyl relations}
  \begin{equation}\label{eq Weyl relations}\begin{split}
  &1.\ \ \ \ W(f)^* = W(-f),\ \ \ \ f \in \BS{H},\\
  &2.\ \ \ \ W(f)W(g) = \exp\big(-i \mbox{Im}\langle f,g\rangle\big)W(f+g), \ \ \ \ f,g \in \BS{H}.
  \end{split}\end{equation}  
It can be shown \cite{Par92} that the Weyl operators 
$W(f)$  $(f\in \BS{H})$ generate the von Neumann algebra of 
all bounded operators on 
$\FC$, i.e.\ $\mbox{vN}\big(W(f);\ f\in \BS{H}\big) = \BC(\FC) =: \WC$. 
It follows from the Weyl relations that for 
all $f \in \BS{H}$ the family $\{W(tf)\}_{t \in \BB{R}}$ forms 
a one-parameter group of unitary operators on $\FC$. It can be shown to be continuous 
in the strong operator topology \cite{Par92}. The following
theorem is a classic result in spectral theory and can 
be found for instance in \cite{KaR83}.

\begin{stel}\label{thm stone}\textbf{(Stone's theorem)}
Let $\{U_t\}_{t\in\BB{R}}$ be a group of unitary operators 
in some von Neumann algebra $\AC$, continuous in the strong 
operator topology. There exists a unique self-adjoint 
operator $A$ affiliated to $\AC$ such that
  \begin{equation*}
  U_t = \exp(itA) := \int_\BB{R}e^{it\lambda}E(d\lambda), 
  \end{equation*}
where $E$ denotes the spectral measure of the self-adjoint operator $A$.  
\end{stel}

Since $\{W(tf)\}_{t\in \BB{R}}$ is a strongly continuous one-parameter
group of unitaries, Stone's theorem provides a self-adjoint 
operator $B(f)$ such that
  \begin{equation*}
  W(tf) = \exp\big(itB(f)\big).
  \end{equation*} 
The operators $B(f)$ $(f\in \BS{H})$ are called 
\emph{field operators}. The domain of the operator $B(f_k)\ldots B(f_1)$
contains $\DC$ for every $f_1,\ldots f_k \in \BS{H}$ and
$k \in \BB{N}$ (cf.\ \cite{Pet90}). For $f,g \in \BS{H}$ and $t \in \BB{R}$
it follows from the Weyl relations that on the domain $\DC$
  \begin{equation}\label{eq prop fields}\begin{split}
  &1.\ \ \   B(tf) = tB(f),          \\
  &2.\ \ \   B(f+g) = B(f) + B(g),          \\
  &3.\ \ \   [B(f), B(g)] = 2 i\mbox{Im}\langle f, g\rangle.
  \end{split}\end{equation}
The last relation (\ref{eq prop fields}.3) is called the 
\emph{canonical commutation relation}. We have introduced 
it via the Weyl operators since they have the advantage 
of being bounded. 

We fix an $\alpha$ in $[0,\pi)$ and denote by $H_\alpha$ the 
subspace of $L^2(\BB{R})$ of functions of the form $e^{i\alpha}f$
with $f$ a real valued function in $L^2(\BB{R})$. 
For $f\in H_\alpha$ we define, as before,  
bounded operators $T(f)_+ := (B(f)+iI)^{-1}$ and 
$T(f)_- := (B(f)-iI)^{-1}$. It follows 
from the canonical commutation relation that 
the family $\{T(f)_+, T(f)_-;\ f\in H_\alpha\}$ 
is commutative and therefore generates a commutative 
von Neumann algebra $\CC_\alpha$. We denote by $\phi$ the 
\emph{vacuum state} on $\WC =\BC(\FC)$, given by 
$\phi(X) := \langle \Phi, X\Phi\rangle$ where  
$\Phi := e(0) = 1\oplus 0 \oplus 0 \oplus \ldots \in \FC$ is 
the \emph{vacuum vector}. Using Theorem \ref{thm spectral}
we obtain a probability space $(\Omega_\alpha, \Sigma_\alpha, {\bf P}_\alpha)$ and a $^*$-isomorphism $\iota:\ \CC_\alpha \to 
L^\infty(\Omega_\alpha,\Sigma_\alpha,{\bf P}_\alpha)$ 
such that $\phi(C) = \BB{E}_{{\bf P}_\alpha}[\iota(C)]$ for all $C \in 
\CC_\alpha$. In a similar fashion as before we can now define 
$\iota(B(f))$ for the (possibly unbounded) self-adjoint operator $B(f)$.   

Let us study the random variable $\iota(B(f))$ on 
the probability space $(\Omega_\alpha, \Sigma_\alpha, {\bf P}_\alpha)$ in 
some more detail by examining its characteristic function 
  \begin{equation}\label{eq char functional}
  \BB{E}_{{\bf P}_\alpha}\Big[\exp\Big(ix\iota\big(B(f)\big)\Big)\Big] = 
  \phi\Big(\exp\big(iB(xf)\big)\Big) = 
  \Big\langle \Phi, W(xf) \Phi\Big\rangle = 
  \exp\Big(-\frac{x^2\p f \p^2}{2}\Big), \ \ \ \ x \in \BB{R}.
  \end{equation}
In the last step we used the definition of the Weyl operator, 
equation \eqref{eq Weyl}, and  the relation 
$\langle e(f), e(g) \rangle = \exp\langle f, g\rangle$, 
which easily follows from equation \eqref{eq exp vector}. 
Define a random process on 
$(\Omega_\alpha,\Sigma_\alpha, {\bf P}_\alpha)$ by
  \begin{equation*} 
  W^\alpha_t := \iota\Big(B\big(e^{i\alpha}\chi_{[0,t]}\big)\Big),\ \ \ \ t\ge 0,  
  \end{equation*}
where $\chi_{[0,t]}$ denotes the indicator function 
of the interval $[0,t]$. From the characteristic functional 
$f \mapsto \BB{E}_{{\bf P}_\alpha}\Big[\exp\Big(i\iota\big(B(f)\big)\Big)\Big]$ 
$(f\in H_\alpha)$ given by 
equation \eqref{eq char functional}, it follows 
that $W_t^\alpha$ is a process with independent, 
normally distributed increments $W^\alpha_t-W^\alpha_s$ 
$(s\le t)$ such that their means are zero and their variances are 
$t-s$. Summarizing, $W^\alpha_t$ 
is a \emph{Wiener process} on 
$(\Omega_\alpha,\Sigma_\alpha, {\bf P}_\alpha)$.   

Identifying the operators $B(f)$ and 
the random variables $\iota(B(f))$, we see that 
the algebra $\WC$ of all bounded operators on the 
Fock space $\FC$ contains a whole family (indexed 
by $\alpha\in [0,\pi)$) of Wiener processes. Note however,
that these processes for different values of $\alpha$, do 
not commute. For example, it follows  
from the cannonical commutation relation 
(\ref{eq prop fields}.3) that $[W^0_s, W^{\pi/2}_t] = 2 i \min\{s,t\} \neq 0$. 
Therefore, for different values of $\alpha$, these processes
can not be represented simultaneously on the same 
probability space via the spectral theorem. It is 
precisely in this sense that noncommutative 
probability is richer than classical probability.

The idea to simultaneously diagonalise the fields in the family
$\{B(e^{i\alpha}\chi_{[0,t]});\ t \ge 0\}$ is implicit is some of the
earliest work in quantum field theory.  However, Segal \cite{Seg56} in the
1950s was the first to emphasise the connection with probability theory.
Apart from the Wiener processes $W^\alpha_t$, the algebra $\WC$ contains
Poisson processes. Before introducing them, some further preparations have
to be made. 

The \emph{second quantisation} of an 
operator $A \in \BC(\BS{H})$ is the operator $\Gamma(A) \in \WC = \BC(\FC)$ 
defined by
  \begin{equation*}
  \Gamma(A) := I \oplus \bigoplus_{n=1}^\infty A^{\ten n}.
  \end{equation*}
For all $A, B \in \BC(\BS{H})$ this immediately gives 
$\Gamma(AB) = \Gamma(A)\Gamma(B)$. Let $S$ be a self-adjoint
element in $\BC(\BS{H})$, then $\exp(itS)$ is a one-parameter
group of unitaries in $\BC(\BS{H})$. After second 
quantisation, this leads to a one-parameter group 
$\Gamma\big(\exp(itS)\big)$ of 
unitaries in $\WC$ (continuous in the strong operator topology). 
Stone's theorem \ref{thm stone} then asserts 
the existense of a self-adjoint operator $\Lambda(S)$ 
on $\FC$ such that for all $t \in \BB{R}$
  \begin{equation*}
  \Gamma\big(\exp(itS)\big) = \exp\big(it\Lambda(S)\big).
  \end{equation*}
The domain of a product $\Lambda(S_1)\ldots\Lambda(S_n)$ contains $\DC$ 
for all self-adjoint
elements $S_1,\ldots, S_n$ in $\BC(\BS{H})$ \cite{Par92}.
Denote by $P_t$ the projection $L^2(\BB{R}) \to L^2(\BB{R}):\ 
f \mapsto \chi_{[0,t]}f$ where $\chi_{[0,t]}$ denotes the 
indicator function of the interval $[0,t]$. For notational convenience, 
we abbreviate the operator $\Lambda(P_t)$ to 
$\Lambda(t)$. On the $n$th-layer of the symmetric Fock space 
$\Gamma(\exp(isP_t))$ acts as $\exp(isP_t)^{\ten n}$. Differentiation 
with respect to $s$ shows that on the $n$th-layer of the 
symmetric Fock space $\Lambda(t) = P_t\ten I^{\ten n-1} + 
I \ten P_t \ten I^{\ten n-2}+\ldots + I^{\ten n-1} \ten P_t$.
This shows that $\Lambda(t)$ is the operator that counts how 
many particles, i.e.\ photons in the field, are present 
in the interval $[0,t]$. In particular we therefore have 
$\Lambda(t)\Phi = 0$, a property that we will exploit 
later on.

Since the family of projections $\{P_t;\ t\ge0\}$ is commutative, it 
generates a commutative von Neumann algebra $\NC$. For all
self-adjoint elements $S$ and $T$ in $\NC$ we have on the 
dense domain $\DC$
  \begin{equation*}\begin{split}
  \big[\Lambda(S),\,\Lambda(T)\big]  = 
  \lim_{t\to 0}\frac{e^{it\Lambda(T)}e^{it\Lambda(S)}
  e^{-it\Lambda(T)}e^{-it\Lambda(S)} - I}{t^2} 
   = \lim_{t\to 0}\frac{\Gamma\big(e^{itT}e^{itS}e^{-itT}e^{-itS}\big)-I}{t^2}
  = 0.
  \end{split}\end{equation*}
For all self-adjoint $S\in \NC$ we can define commuting bounded operators
$T(S)_+ := (\Lambda(S)+iI)^{-1}$ and $T(S)_- := (\Lambda(S)-iI)^{-1}$. 
The above ensures that they generate a 
commutative von Neumann subalgebra $\CC$ of $\WC$. 
For $f\in L^2(\BB{R})$ we define a \emph{coherent vector} $\psi(f)$ by 
  \begin{equation*}
  \psi(f) := W(f)\Phi = W(f)e(0) = \exp\Big(-\frac{\p f\p^2}{2}\Big)e(f). 
  \end{equation*}
A \emph{coherent state} $\rho$ on $\WC$ is defined  
by $\rho(A) := \langle\psi(f), A\psi(f)\rangle$. 
The spectral theorem \ref{thm spectral} provides  
a classical probability space $(\Omega,\Sigma,{\bf P})$ 
and a $^*$-isomorphism $\iota:\ \CC\to L^\infty(\Omega,\Sigma,{\bf P})$ 
such that $\rho(C) = \BB{E}_{{\bf P}}[\iota(C)]$. For all
self-adjoint operators $S$ in the von Neumann algebra $\NC$, 
we can define as before random variables $\iota\big(\Lambda(S)\big)$ 
on $(\Omega,\Sigma,{\bf P})$.   

Let us investigate the characteristic function of the 
random variable $\iota\big(\Lambda(S)\big)$ in some 
more detail
  \begin{equation*}\begin{split}
  \BB{E}_{{\bf P}}\Big[e^{ix\iota\big(\Lambda(S)\big)}\Big] & = 
  \Big\langle\psi(f), e^{ix\Lambda(S)}\psi(f)\Big\rangle =
  \Big\langle\psi(f), \Gamma\big(e^{ixS}\big)\psi(f)\Big\rangle \\ 
  & = e^{-\p f\p^2}\Big\langle e(f), e\big(e^{ixS}f\big)\Big\rangle 
   = e^{\big\langle f, \big(e^{ixS}-I\big)f\big\rangle}.
  \end{split}\end{equation*}
The functional 
$S \mapsto \BB{E}_{{\bf P}}\Big[e^{i\iota\big(\Lambda(S)\big)}\Big]$  
shows that the process $N_t:=\iota\big(\Lambda(t)\big)$ 
is a Poisson process \cite{HuP84} on $(\Omega, \Sigma, {\bf P})$ 
with intensity measure $|f|^2d\lambda$, where $\lambda$ stands for 
the Lebesgue measure. Summarizing, when photons are 
counted in a laser beam they arrive Poisson distributed. 
Since $\psi(f) = W(f)e(0) = W(f)\Phi$, we could just as well 
have studied the commutative von Neumann algebra $W(f)^*\CC W(f)$, 
equipped with the vacuum state $\phi$. In this way we 
get a whole family of Poisson processes indexed by $f$ 
within the quantum probability space $(\WC,\phi)$. 
Again, these processes do not commute amongst each other \cite{HuP84}.

In this section we have seen that the quantum 
probability space $(\WC,\phi)$ contains many 
interesting classical stochastic processes. 
However, these classical processes 
do not commute amongst each other. 
We will proceed by discussing the stochastic calculus 
of Hudson and Parthasarathy \cite{HuP84} enabling 
us to treat the stochastic analysis of all these processes 
in one framework.

\section{Quantum stochastic calculus}\label{sec qsc}

We start this section with some technical definitions and manipulations to
clear the way for defining the stochastic integrals of Hudson and
Parthasarathy. However, it is not so much the definition of the stochastic
integrals that is of the greatest importance here.  It is the subsequent
It\^o rule obeyed by these stochastic integrals, summarized in their It\^o
table, that will enable us to put them to good use. The It\^o rule
translates the difficult analysis involved in defining the stochastic
integrals into simple algebraic manipulations with increments. That is the
real strength of having a stochastic calculus. We refer to
\cite{HuP84,Bia95,Mey93,Par92} for more extensive treatments of quantum
stochastic calculus. 

Let $\BS{H}_1$ and $\BS{H}_2$ be Hilbert 
spaces. The symmetric Fock space has the 
following \emph{exponential property} (cf.\ \cite{Par92})
  \begin{equation*}
  \FC(\BS{H}_1\oplus\BS{H}_2) \cong \FC(\BS{H}_1) \ten \FC(\BS{H}_2),
  \end{equation*}
in the sense that there exists a unitary operator 
$U:\ \FC(\BS{H}_1\oplus\BS{H}_2)\to \FC(\BS{H}_1)\ten\FC(\BS{H}_2)$ 
such that for all $f_1 \in \BS{H}_1$ and $f_2\in \BS{H}_2$ 
the exponential vector $e(f_1\oplus f_2)$ is mapped to the tensor product 
$e(f_1)\ten e(f_2)$. For $s\le t$ and $f\in L^2(\BB{R})$ 
we write $f_{t]} := \chi_{(-\infty,t]}f$, 
$f_{[t} := \chi_{[t,\infty)}f$ and $f_{[s,t]} := \chi_{[s,t]}f$. Furthermore we have 
$L^2(\BB{R}) = L^2\big((-\infty,t]\big)\oplus L^2\big([t,\infty)\big)$, 
which means that every $f$ in $L^2(\BB{R})$ can be 
uniquely written as a sum $f = f_{t]} + f_{[t}$ of elements 
in $L^2\big((-\infty,t]\big)$ and $L^2\big([t,\infty)\big)$. 
Writing $\FC_{t]} := \FC\big(L^2\big((-\infty,t]\big)\big)$, 
$\FC_{[t}:= \FC\big(L^2\big([t,\infty)\big)\big)$ and $\FC_{[s,t]}:=\FC\big(L^2([s,t])\big)$, 
this 
leads to the splitting $\FC = \FC_{t]}\ten\FC_{[t}$ where, as before,  
exponential vectors of direct sums are identified with tensor 
products of exponential vectors. Since $t$ can vary continuously 
through $\BB{R}$, $\FC$ is said to be a \emph{continuous tensor product}.
The algebra of all bounded operators on $\FC$ splits in a similar 
way, that is $\WC = \BC(\FC) = \BC(\FC_{t]})\ten\BC(\FC_{[t}) = 
\WC_{t]}\ten\WC_{[t}$ (cf.\ \cite{Par92}), where we denote 
$\WC_{t]} := \BC(\FC_{t]})$, $\WC_{[t} := \BC(\FC_{[t})$ and
$\WC_{[s,t]} := \BC(\FC_{[s,t]})$. 

On the domain $\DC$ we introduce \emph{annihilation operators} 
$A_t$ and \emph{creation operators} 
$A^*_t$  by 
  \begin{equation*}
  A_t := \frac{1}{2}\big(B(i\chi_{[0,t]}) - iB(\chi_{[0,t]})\big)\ \ \ \
  \mbox{and} \ \ \ \
  A^*_t := \frac{1}{2}\big(B(i\chi_{[0,t]}) + iB(\chi_{[0,t]})\big). 
  \end{equation*}
It can be shown \cite{Par92} that $A_te(f) = \langle \chi_{[0,t]}, f\rangle e(f)$ 
and that $A^*_t$ is its adjoint on $\DC$. This means that these operators are the  
annihilation and creation operators for the mode $\chi_{[0,t]}$ of 
the field as they are known in physics. Note that $A_t\Phi = A_te(0) = 0$, 
a property that we will exploit in future.
We will denote by $\Lambda_t$ the restriction of 
$\Lambda(t)$ to the domain $\DC$. It can be shown \cite{Par92} 
that $\langle e(g), \Lambda_t e(f)\rangle = 
\langle g, \chi_{[0,t]}f\rangle \langle e(g), e(f)\rangle$.
This implies the important relation $\Lambda_t\Phi=0$ that 
we already encountered in the previous section. 

Let $M_t$ be one of the processes $A_t, A^*_t$ or $\Lambda_t$. 
The following \emph{factorisation property} \cite{HuP84,Par92} 
underlies the definition of the stochastic integral
  \begin{equation*}
  (M_t-M_s) e(f) = e(f_{s]})\Big((M_t-M_s)e(f_{[s,t]})\Big)e(f_{[t}),\ \ \ \ s\le t,  
  \end{equation*}
such that $(M_t-M_s)e(f_{[s,t]})\in \FC_{[s,t]}$. We have made our 
notation lighter by omitting the tensor product signs. 
Let $\HC$ be a Hilbert space, called the \emph{initial 
space}. 
We tensor the initial space to $\FC$ and extend the 
operators $A_t$, $A^*_t$ and $\Lambda_t$ to $\HC\ten\FC$ 
by ampliation, i.e.\ by tensoring the identity to them on 
$\HC$ (however, to keep notation light we will not denote it). 
Just for mathematical convenience we will take $\HC$ to be finite 
dimensional, i.e.\ $\HC = \BB{C}^n$. We denote the algebra 
of all operators on $\HC$ by $\BC$. We are ready for the 
definition of the stochastic integral. 

\begin{de}\label{def stoch int}\textbf{(Quantum stochastic integral)} 
Let $\{L_s\}_{0 \le s \le t}$ be an adapted 
(i.e.\ $L_s \in \BC \ten \WC_{s]}$ for all $0 \le s \le t$) simple 
process with respect to the partition $\{s_0=0, s_1,\dots, s_p= t\}$ in the sense that 
$L_s = L_{s_j}$ whenever $s_j \le s < s_{j+1}$. The stochastic integral of 
$L$ with respect to $M$ on $\BB{C}^n \ten \DC$ is then defined as 
\cite{HuP84,Par92}  
  \begin{equation*}
  \int_0^t L_s dM_s  ~x e(f) := 
\sum_{j=0}^{p-1} \Big(L_{s_j}xe(f_{s_j]})\Big)\Big((M_{s_{j+1}}- M_{s_{j}})
   e(f_{[s_j, s_{j+1}]})\Big)e(f_{[s_{j+1}}),\ \ \ \ x \in \BB{C}^n.
  \end{equation*}
The notation is simplified by writing $dX_t = L_tdM_t$ for $X_t = X_0 + 
\int_0^t L_sdM_s$. 
The definition of the stochastic integral can be extended to a large class
of {stochastically integrable processes} \cite{HuP84,Par92} if we 
approximate these by simple functions and take a limit in the strong 
operator topology.  
\end{de}

Since $A_t\Phi =\Lambda_t\Phi =0$ it is immediate from the 
definition that quantum stochastic integrals with respect 
to $A_t$ and $\Lambda_t$ acting on $\Phi$ are zero, or 
infinitesimally $d\Lambda_t\Phi_{[t} = dA_t\Phi_{[t} =0$. From 
this we can immediately conclude that vacuum expectations 
of stochastic integrals with respect to $A_t$ and 
$\Lambda_t$ vanish. Furthermore, 
we have $\big\langle \Phi, \int_0^t L_sdA^*_s\Phi\big\rangle = 
\big\langle\int_0^t L^*_sdA_s \Phi, \Phi\big\rangle = 0$, i.e.\ vacuum 
expectation of stochastic integrals with respect to $A^*_t$ are 
zero as well. Note, however, that $dA^*_t\Phi_{[t} \neq 0$. 

To get some more feeling for the definition of the 
quantum stochastic integral we will now investigate which 
quantum stochastic differential equation is satisfied by 
the Weyl operators $W(f_{t]})$ $(f\in L^2(\BB{R}))$. 
Note that the stochastic integrals are 
defined on the domain $\DC$. Therefore we calculate
for $g$ and $h$ in $L^2(\BB{R})$ 
  \begin{equation*}
  \phi(t) := \big\langle e(g) , W(f_{t]})e(h)\big\rangle = 
  e^{-\langle f_{t]}, h\rangle-\frac{1}{2}\p f_{t]}\p^2} 
  \big\langle e(g) , e(h+f_{t]})\big\rangle = 
  e^{\langle g, f_{t]}\rangle -\langle f_{t]}, h\rangle-\frac{1}{2}\p f_{t]}\p^2}e^{\langle g,h\rangle},
  \end{equation*}
which means that
  \begin{equation*}
  \phi(t)-\phi(0) = \int_0^t\Big\langle e(g),
  \frac{d}{ds}\Big(\langle g, f_{s]}\rangle -\langle f_{s]}, h\rangle
  -\frac{1}{2}\p f_{s]}\p^2\Big)\phi(s) e(h)\Big\rangle ds.   
  \end{equation*}
Let us turn to the definition of the stochastic 
integral, Definition \ref{def stoch int}. Let 
$\{0=s_0, s_1,\ldots, s_p =t\}$ be a partition of $[0,t]$ and 
choose $L_s = \overline{f}(s_j)W(f_{s_j]})$ for $s_j \le s < s_{j+1}$. Let further 
$M_t$ be $A_t$, then the definition of the stochastic integral 
gives (heuristically in the last step)
  \begin{equation*}\begin{split}
  &\sum_{j=0}^{p-1} \Big(\overline{f}(s_j)W(f_{s_j]})e(h_{s_j]})\Big)\Big((A_{s_{j+1}}- A_{s_{j}})
   e(h_{[s_j, s_{j+1}]})\Big)e(h_{[s_{j+1}}) = \\
  &\sum_{j=0}^{p-1} \Big\langle f(s_j) \chi_{[s_j,s_{j+1}]}, h\Big\rangle W(f_{s_j]})e(h) = 
  \sum_{j=0}^{p-1} 
\Big(\big\langle f(s_j) \chi_{s_{j+1}]}, h\big\rangle-
\big\langle f(s_j) \chi_{s_{j}]}, h\big\rangle\Big)W(f_{s_j]})e(h)  \\  
  &\longrightarrow \int_0^t d\big\langle f_{s]}, h\big\rangle\,W(f_{s]})\,e(h).
  \end{split}\end{equation*}
Together with a similar calculation for $M_t =A_t^*$, this yields 
the following quantum stochastic differential equation for the 
Weyl operator $W(f_{t]})$ 
  \begin{equation}\label{eq dif Weyl}
  dW(f_{t]}) = \Big\{f(t)dA^*_t-\overline{f}(t)dA_t - \frac{1}{2}\big|f(t)\big|^2dt\Big\}W(f_{t]}).
  \end{equation}

Let us return to developing the theory further. To 
be able to compute with the stochastic integral, we 
need a stochastic calculus, i.e.\ a quantum It\^o 
rule. The basic ingredient for its proof comes from 
the commutation relations between $A_t, A_t^*$ and 
$\Lambda_t$. For instance, from the canonical  
commutation relation and the definition of $A_t$
and $A_t^*$ we see $[A_t, A_t^*] = t$ on the 
exponential domain $\DC$. Therefore we have
  \begin{equation*}
  \big\langle e(f), A_tA_t^*e(g)\big\rangle =  
  \big\langle A_te(f), A_t e(g)\big\rangle + t\big\langle e(f),e(g)\big\rangle = 
  \Big(\big\langle f,\chi_{[0,t]}\big\rangle  \big\langle \chi_{[0,t]}, g\big\rangle +t\Big)
  \big\langle e(f),e(g)\big\rangle.
  \end{equation*}
Infinitesimally this immediately leads to 
$d(A^*_tA_t) = A^*_tdA_t+A_tdA_t^* + dt$. The next 
theorem is built on similar ideas, it can be found 
in $\cite{HuP84}$ and $\cite{Par92}$.

\begin{stel}\label{Itorule}\textbf{(Quantum It\^o rule \cite{HuP84})}
Let $X_t$ and $Y_t$ be stochastic integrals of the form 
\begin{equation*}\begin{split}
&dX_t = B_td\Lambda_t + C_t dA_t + D_t dA_t^* + E_tdt, \\
&dY_t = F_t d\Lambda_t + G_tdA_t + H_t dA_t^* + I_tdt,
\end{split}\end{equation*}
for some stochastically integrable processes $B_t, C_t, D_t, E_t, F_t, 
G_t, H_t$ and $I_t$ (see \cite{HuP84,Par92}). 
The process $X_tY_t$ then satisfies the relation
  \begin{equation*}
  d(X_tY_t) = X_t\,dY_t + (dX_t)\,Y_t + dX_t\,dY_t,
  \end{equation*}
on the domain $\mathcal{D}$, where $dX_tdY_t$ should be evaluated 
according to the quantum It\^o table
\begin{center}
{\large \begin{tabular} {l|llll}
 & $dA_t$  & $d\Lambda_t$ & $dA^*_t$ & $dt$\\
\hline 
$dA_t$ & $0$ & $dA_t$ & $dt$ & $0$\\
$d\Lambda_t$ & $0$ & $d\Lambda_t$ & $dA^*_t$ & $0$ \\
$dA^*_t$ & $0$ & $0$  & $0$ & $0$\\
$dt$ & $0$ & $0$ & $0$ & $0$
\end{tabular} }
\end{center}
i.e.\ $dX_tdY_t = B_tF_td\Lambda_t + C_tF_tdA_t + B_tH_tdA^*_t + C_tH_tdt$.
\end{stel}

In the previous section we encountered the classical 
Wiener processes $B\big(e^{i\alpha}\chi_{[0,t]}\big) = 
ie^{-i\alpha}A_t -ie^{i\alpha}A^*_t$ for $\alpha\in [0,\pi)$. Since 
$dB\big(e^{i\alpha}\chi_{[0,t]}\big) = 
ie^{-i\alpha}dA_t -ie^{i\alpha}dA^*_t$, we 
recover the classical It\^o rule for these Wiener 
processes, i.e.\  $\big(dB\big(e^{i\alpha}\chi_{[0,t]}\big)\big)^2 = dt$, 
from the quantum It\^o rule, as it should. For an $f\in L^2(\BB{R})$
we can write the Weyl operator $W(f_{t]})$ as  
$W(f_{t]}) = \exp\big(\int_0^t f(s)d(A^*_s-A_s)\big)$. 
Therefore it follows from the quantum It\^o rule that 
the Weyl operators $W(f_{t]})$ satisfy equation \eqref{eq dif Weyl}, 
where $-\frac{1}{2}\p f\p^2W(f_{t]})dt$ is the It\^o correction 
term. The Poisson process of the previous section was 
given by $\Lambda^f_t := W(f)^*\Lambda_tW(f)$ in the 
vacuum state, for which 
the quantum It\^o rule gives
  \begin{equation*}
  d\Lambda^f_t = d\Lambda_t + \overline{f}(t)dA_t + f(t)dA_t^* + |f(t)|^2dt, 
  \end{equation*}
which leads to the classical It\^o rule $(d\Lambda^f_t)^2= d\Lambda^f_t$ 
for the Poisson process. Furthermore, integrating the above equation, 
we see that $\Lambda^f_t = \Lambda_t + B(if_{t]}) + \int_0^t|f(s)|^2ds$.

In the next section, we will show that quantum stochastic differential 
equations (QSDEs) of the following form 
  \begin{equation}\label{eq Ut}
  dU_t = \Big\{LdA_t^* -L^*dA_t - \frac{1}{2}L^*Ldt - iHdt\Big\}U_t, \ \ \ \ U_0 = I,
  \end{equation}
where $L$ and $H$ are in $\BC$ and $H$ is self-adjoint, emerge naturally 
from physical models.  A standard Picard iteration argument 
\cite{HuP84,Par92} ensures existence and uniqueness of the solution.   
The adjoint $U^*_t$ satisfies 
  \begin{equation*}
  dU^*_t = U^*_t\Big\{L^*dA_t -LdA^*_t - \frac{1}{2}L^*Ldt + iHdt\Big\}, \ \ \ \ U^*_0 = I.
  \end{equation*}
From the quantum It\^o rule it now immediately follows that $d(U^*_tU_t) = 0$, 
which means that the solution $U_t$ is unitary for all $t$.  The 
interpretation is that $U_t$ defines a time evolution or {\it flow} 
$X\mapsto U_t^*XU_t$ in $\mathcal{B}\otimes\mathcal{W}$, 
i.e., an observation of $X$ at time $t$ is described by the observable 
$U_t^*XU_t$.

{\bf Remark.}
Despite the fact that both $A_t+A_t^*$ and $i(A_t-A_t^*)$ are classical
noises, Eq.\ (\ref{eq Ut}) is not a classical stochastic 
differential equation.  To make this idea explicit, rewrite the equation as
\begin{equation}\label{eq HP again}
  dU_t = \Big\{iL_+i(dA_t-dA_t^*) + iL_-(dA_t+dA_t^*) - 
  \frac{1}{2}L^*Ldt - iHdt\Big\}U_t, \ \ \ \ U_0 = I,
\end{equation}
where $L_+ = (L+L^*)/2$ and $L_- = (L-L^*)/(2i)$. 
This reveals particularly well that the initial system is driven 
simultaneously by two noncommuting noises.  If either $L_+=0$ or $L_-=0$, 
we say that $U_t$ is {\it essentially commutative} \cite{KuM87}.

Define $T_t(X) : = \mbox{id}\ten\phi({U}_t^*\,X\ten I\,U_t)$ for 
all $X \in \BC$ and $t\ge 0$. Using the quantum It\^o rule and the fact 
that vacuum expectations of stochastic integrals vanish, we find for all 
$X \in \BC$
  \begin{equation*}
  dT_t(X) = \mbox{id}\ten\phi\big(d(U_t^*X\ten I U_t)\big) = 
  T_t\big(\mathcal{L}_{L,H}(X)\big)dt,
  \end{equation*}  
where we have defined the Lindblad generator \cite{Lin76}
  \begin{equation*}
  \mathcal{L}_{L,H}(X) = i[H,X] + L^*XL - 
  \frac{1}{2}(L^*LX+XL^*L), \ \ \ \ X \in \BC.
  \end{equation*}   
In quantum probability, this object plays the same role as the 
infinitesimal generator of a Markov diffusion in classical probability 
theory.  Indeed, the semigroup $T_t$, describing the time evolution of 
the expectation of any system observable, can be written as $T_t = 
\exp(t\mathcal{L}_{L,H})$.  The commutator with $H$ describes the unitary 
evolution of the initial system itself, whereas the residual terms describe 
the irreversible effect that the interaction with the environment 
has on the initial system. 

A more general case can be treated in a similar fashion by introducing 
more channels in the field. That is, the initial system $\BB{C}^n$ is 
tensored to $k$ Fock spaces. The QSDE that defines the time evolution of the 
initial system and the field together is given in a general form by
  \begin{equation*}
  dU_t = \big\{L_jdA_j^*(t) + (S_{ij}- \delta_{ij})d\Lambda_{ij}(t) - 
  L^*_iS_{ij}dA_j(t) -(iH + \frac{1}{2} L_j^*L_j)dt\big\}U_t,\ \ \ \ U_0 =I,
  \end{equation*}
where repeated indices are being summed. The index $j$
on $A_j(t)$ and $A_j(t)^*$ labels the annihilator and creator 
on the $j$th copy of the Fock space, $S$ is a unitary operator 
on $\BB{C}^n\ten l^2(\{1,2,\ldots,k\})$ such that 
$S_{ij} = \langle i, S j\rangle$, and $\Lambda_{ij}(t) := 
\Lambda(P_t\ten|i\rangle\langle j|)$. Physically, the term 
$(S_{ij}-\delta_{ij})d\Lambda_{ij}(t)$ describes direct scattering between 
channels $i$ and $j$ in the field \cite{BaL00}. For this equation
we again have existence, uniqueness and unitarity of its solution \cite{Par92}. 
Moreover, we obtain the associated generator
  \begin{equation}\label{eq genLin}
  \mathcal{L}(X) = i[H,X] + \sum_{j=1}^k L_j^*XL_j - 
  \frac{1}{2}\{L_j^*L_j, X\}, \ \ \ \ X \in \BC,
  \end{equation}
where $\{X,Y\}$ stands for the anti-commutator $XY+YX$.  It was shown by 
Lindblad \cite{Lin76} that any semigroup of completely positive identity 
preserving operators on $\mathcal{B}=M_n(\mathbb{C})$ has a generator of 
the form Eq.\ (\ref{eq genLin}).

\section{The filtering problem in quantum optics}\label{sec filtering}

Beside the intrinsic interest of quantum probability as a mathematical 
generalization of classical probability theory, many realistic physical 
scenarios are very well described by QSDEs of the form we have discussed.  
The inherent stochasticity of quantum mechanical problems makes this 
field a rich playground for the application of the classical theories of 
statistical inference and control of stochastic processes.  Of course, as 
in the classical theory, white noise systems are only an idealization of 
physical interactions; a Markov limit of wide-band noise in the spirit of 
Wong and Zakai gives stochastic models in the It\^o form.  For a large 
class of quantum systems, particularly those arising in the field of 
quantum optics, such approximations are extremely good and describe 
laboratory experiments essentially to experimental precision.  Though a 
detailed discussion of the physics involved in the modelling of such 
systems is beyond the scope of these notes, we here briefly describe the 
physical origin of the equations that are widely used in the physics 
community \cite{GC84}.

The basic model of quantum optics consists of some fixed physical system,
e.g.\ a collection of atoms, in interaction with the electromagnetic
field. The atomic observables are self-adjoint operators in a Hilbert
space $\mathcal{H}$.  The description of the electromagnetic field and its
interaction with the atoms follows from basic physical arguments (e.g.\
quantization of Maxwell's equations; see the excellent monograph
\cite{CT89} for a thorough treatment of this field, known as {\it quantum 
electrodynamics}.)  It turns out that the free electromagnetic field, 
i.e.\ an optical field in empty space, is described by a stationary Gaussian 
(noncommutative) wide band noise $\tilde a(t,{\bf r})$ that propagates 
through space at the speed of light $c$; i.e.\ if we restrict ourselves 
to a single spatial dimension, $\tilde a(t+\tau,z)=\tilde a(t,z-c\tau)$.  
If we now place the atoms at the origin $z=0$, then the quantum dynamics is 
given by a Schr\"odinger equation of the form
$$
	\frac{d}{dt}\tilde U(t)=
	-\frac{i}{\hbar}\left[H-i\hbar(L(t)+L^*(t))(\tilde a(t,0)-
		\tilde a^*(t,0))
	\right]\tilde U(t),
$$
where $L(t)$ is an atomic (dipole) operator and $H$ is an atomic 
Hamiltonian.  This equation, which follows directly from the physical 
model, is similar to (\ref{eq Ut}) but has wide-band right hand side.  
One generally assumes that the dipole operator is harmonic,
$L(t)=Le^{-i\omega t}$, but this assumption can be relaxed \cite{AGLu95}.
Hence $L$ and $H$ are operators in the initial system $\BC$, $H$ being 
self-adjoint. 

We now wish to approximate this equation by an equation where the noise 
has infinite bandwidth.  There is a large body of 
literature on such approximations in classical probability, following the
pioneering work of Wong and Zakai (see e.g.\ \cite{Kush79} and references 
therein.)  A common way to attain this limit is to rescale the 
(wide-band) noise term by $1/\varepsilon$, time as $t/\varepsilon^2$, 
and then take the limit $\varepsilon\to 0$.  The effect of this rescaling 
\cite{Kush80} is that the noise bandwidth goes to infinity, whereas its 
energy per unit bandwidth is retained.  A similar intuition holds in the 
noncommutative case; i.e., one can show that for the type of model we 
have described
$$
	\frac{1}{\varepsilon}
	\int_0^t L(s/\varepsilon^2)(\tilde a^*(s/\varepsilon^2,0)
		-\tilde a(s/\varepsilon^2,0))\,ds\longrightarrow
	\sqrt{\gamma}\,LA_t^*\mbox{~~as~~}\varepsilon\longrightarrow 0
	\mbox{~~(in distribution),}
$$
where $\gamma>0$ depends on the characteristics of the wide-band noise
$\tilde a$ and on the dipole rotation frequency $\omega$ 
\cite{AFLu90,AGLu95,Gou05}.  

The essential question is now the limiting 
behavior as $\varepsilon\to 0$ of the rescaled Schr\"odinger equation
\begin{equation}
\label{eq:vanhove}
	\frac{d}{dt}\tilde U^\varepsilon(t)=
	-\frac{i}{\hbar}\left[H-i\hbar\left(L(t/\varepsilon^2)+
		L^*(t/\varepsilon^2)\right)\left(\tilde a(t/\varepsilon^2,0)-
		\tilde a^*(t/\varepsilon^2,0)\right)/\varepsilon
	\right]\tilde U^\varepsilon(t).
\end{equation}
The limit $\varepsilon\to 0$ of this equation is known as the {\it weak 
coupling limit}, or the Friedrichs-Van Hove limit, in the quantum 
probability literature.  The limit can be studied by expanding the 
solution of Eq.\ (\ref{eq:vanhove}) in a (Dyson) series using Picard 
iteration, and then calculating the limit of each term in the series 
\cite{AFLu90,AGLu95,Gou05,Gou04}.  In particular, as we are interested in 
the time evolution $\tilde U(t)^*X(t)\tilde U(t)$ of any observable $X$, one 
can study the convergence of each term in the series expansion of $\tilde 
U(t)^{\varepsilon*}X\tilde U^\varepsilon(t)$.  One finds that
$\tilde U(t)^{\varepsilon*}X\tilde U^\varepsilon(t)\to U_t^*XU_t$ as 
$\varepsilon\to 0$, where 
$$
  dU_t = \Big\{\sqrt{\gamma}\,LdA_t^*-\sqrt{\gamma}\,L^*dA_t - 
\frac{1}{2}(\gamma+i\sigma)L^*Ldt - iHdt\Big\}U_t, \ \ \ \ U_0 = I, 
$$
with $\sigma\in\mathbb{R}$.  The additional term $\propto L^*L$ is 
the It\^o correction term obtained in the limit.  It is a signature of the 
noncommutativity of the wide band noise that the constant multiplying the 
It\^o correction is complex; the correction consists of a diffusive term 
$\gamma L^*L/2$ and a Hamiltonian term $i\sigma L^*L/2$.  In the 
following we will absorb the constant $\gamma$ into the 
definition of $L$, i.e.\ $\sqrt{\gamma}\,L\mapsto L$, and the Hamiltonian 
correction into the definition of $H$, i.e.\ $H+\sigma L^*L/2\mapsto H$.
This gives 
\begin{equation}\label{eq HP}
  dU_t = \Big\{LdA_t^*-L^*dA_t - 
	\frac{1}{2}L^*Ldt - iHdt\Big\}U_t, \ \ \ \ U_0 = I, 
\end{equation} 
which is a QSDE in the Hudson-Parthasarathy form like we encountered 
previously.  Henceforth we will take this equation as our physical model.  
We note that in some situations it is also possible to obtain gauge 
($d\Lambda_t$) processes in the weak coupling limit \cite{Gou05}; this is 
particularly useful for the description of a strong off-resonant laser 
probe, which is essentially a scattering interaction.

In classical models, the system dynamics is usually described by an SDE 
which determines the time evolution of the state of the system.  A 
classical ``system observable'', then, is any function $f(x_t)$ of the 
system state $x_t$.  The flow $j_t$ of the dynamics then returns the 
random variable $j_t(f)=f(x_t)$ corresponding to the observable $f$ at 
time $t$.  Due to the fact that quantum models are noncommutative we 
cannot give a sample path interpretation to the full model; in 
particular, there is no analogue of the system state $x_t$ in quantum 
mechanics.  On the other hand, we do have a well-defined concept of a 
system observable, which is just any observable of the initial system.
The unitary solution $U_t$ of Eq.\ (\ref{eq HP}) then provides a quantum 
analogue of the flow $j_t$, given by $j_t(X)=U_t^* XU_t$.  This description 
of a stochastic model by a fixed probability measure (state) and 
time-varying random variables (observables), which is nearly universal in 
classical stochastics, is known as the {\it Heisenberg picture} in 
quantum mechanics.  It is not difficult, using the It\^o rules, to find a 
dynamical equation for $j_t$
\begin{equation}
	dj_t(X)=j_t(\mathcal{L}_{L,H}(X))\,dt+
		j_t([L^*,X])\,dA_t+j_t([X,L])\,dA_t^*.
\end{equation}
This is the quantum analogue of the classical It\^o formula for $j_t(f)$.

Having described the dynamics of the system and its interaction with the 
field, let us now turn to the observations that we can perform.  Unlike 
in classical stochastic theory, where one observes some observable of the 
system, in quantum models an observation is generally performed in the 
field.  From the system's perspective, the interaction with the field 
looks like an (albeit noncommutative) noisy driving force.  Similarly, 
however, the field is perturbed by its interaction with the atoms, and 
carries off information as it propagates away after the interaction.  By 
performing a measurement in the field, then, we can attempt to perfom 
statistical inference of the atomic observables.

To calculate the perturbation of the field by the atoms we once again
calculate $U_t^*YU_t$, where now, however, $Y$ is a field observable. The
field observable of interest depends on the type of measurement we choose
to perform.  Without entering into the details, we mention two types of 
measurement that are extremely common in quantum optics:  direct 
photodetection (photon counting), for which the observation at time 
$t$ is given by $Y_t^\Lambda=U_t^*\Lambda_tU_t$, and homodyne detection, for 
which $Y_t^W=U_t^*(A_t+A_t^*)U_t$ (more generally
$Y_t^W=U_t^*(e^{-i\varphi}A_t+e^{i\varphi}A_t^*)U_t$.)  We refer to 
\cite{Bar90,Bar03} for a detailed treatment of quantum optical 
measurements.  Using the It\^o rules we obtain 
$$
	dY_t^\Lambda=d\Lambda_t+j_t(L)\,dA_t^*+j_t(L^*)\,dA_t+j_t(L^*L)\,dt,
$$
$$
	dY_t^W=j_t(L+L^*)\,dt+dA_t+dA_t^*.
$$
Intuitively, it appears that $Y_t^\Lambda$ is like a Poisson process whose
intensity is controlled by $j_t(L^*L)$, whereas $Y_t^W$ is a noisy
observation of $j_t(L+L^*)$.  We cannot draw this conclusion, however, as
$j_t(L)$ need not commute with $A_t$ or $A_t^*$, nor with itself at
different times. 

It is essential, however, that the observation process commutes with 
itself at different times, and is hence equivalent to a classical 
stochastic process through the spectral theorem.  An observation process 
that does not obey this property cannot be observed in a single realization 
of an experiment and is physically meaningless.  Let us show that the 
observations processes we have defined above do obey this property, 
which is called the {\it self-nondemolition} property.  Let $Z$ be any 
operator of the form ${I}\otimes Z_{s]}\otimes{I}$ on
$\mathcal{H}\otimes\mathcal{F}_{s]}\otimes\mathcal{F}_{[s}$ and let $t\ge 
s$.  Then the It\^o rules give directly
$$
	U_t^*ZU_t=U_s^*ZU_s+\int_s^t
	U_\tau^*\mathcal{L}_{L,H}(Z)U_\tau\,d\tau
	+\int_s^t U_\tau^*[L^*,Z]U_\tau\,dA_\tau
	+\int_s^t U_\tau^*[Z,L]U_\tau\,dA_\tau^*.
$$
Now let $Z=A_s+A_s^*$ or $Z=\Lambda_s$.  In both cases 
$\mathcal{L}_{L,H}(Z)=[Z,L]=0$ as $L$ and $H$ are system observables and 
$Z$ is a field observable.  Hence $Y_s^W=U_t^*(A_s+A_s^*)U_t$ and 
$Y_s^\Lambda=U_t^*\Lambda_s U_t$ for all $t\ge s$.  It is now easily 
verified, using the unitarity of $U_t$ and the fact that $A_s+A_s^*$ and 
$\Lambda_s$ are commutative processes, that 
$[Y_t^W,Y_s^W]=[Y_t^\Lambda,Y_s^\Lambda]=0$ for all $t,s$.  Do note, 
however, that $Y_t^W$ and $Y_t^\Lambda$ do not commute with each other; 
in any experimental realization, we can choose to perform only one of these 
measurements.

Moving on to the next step in our program, we now wish to use the
information gained from the measurement process to infer something about
the initial system.  To find a least mean square estimate of a system
observable $X$ at time $t$, given the observations $Y_t$ up to this time,
we must calculate $\pi_t(X)= \mathbb{P}(j_t(X)|\mathcal{Y}_t)$, where
$\mathcal{Y}_t=\mathrm{vN}(Y_s:0\le s\le t)$ is the (commutative) Von
Neumann algebra generated by the observations history.  The remainder of
these notes is devoted to finding a recursive equation for $\pi_t(X)$ (the
{\it filtering equation}).  As we have discussed previously, however, this
conditional expectation is only defined if $j_t(X)$ is in the commutant of
$\mathcal{Y}_t$, the intuitive interpretation being that statistical
inference of an observable is only physically meaningful if the
conditional statistics could possibly be tested through a compatible
experiment.  Through an entirely identical procedure as in the previous
paragraph, we can show that $j_t(X)$ is in the commutant of
$\mathcal{Y}_t$ for any system observable $X$.  This is known as the {\it
nondemolition property}. 

{\bf Remark.} Unlike in a classical filtering scenario, we have not added
any independent corrupting noise to the observations.  Nonetheless, the
filtering problem does not trivialize to a problem with complete
observations because the system is driven by a quadrature of the field
that does not commute with the observations.  Hence the problem of partial
observations is intrinsic to quantum measurement theory, and does not
result only due to technical noise in the detection apparatus.  Additional
technical noise is however straightforward to take into account as well;
we will discuss this possibility in a later section. 

As a final note, we remark that we have now obtained a system-theoretic
model of our system and observations.  For example, the algebra 
$(\mathcal{B}\otimes\mathcal{W},\rho\otimes\phi)$ together with the pair
$$
	dj_t(X)=j_t(\mathcal{L}_{L,H}(X))\,dt+
		j_t([L^*,X])\,dA_t+j_t([X,L])\,dA_t^*
$$
$$
	dY_t=j_t(L+L^*)\,dt+dA_t+dA_t^*
$$
completely defines a system-observations model, in direct analogy to the 
system-observation models used throughout classical nonlinear filtering 
and stochastic control theory.

\section{The reference probability method}\label{sec KS formula}

The goal of this section is to derive the quantum filtering equation of 
Belavkin, a recursive equation for $\pi_t(X)$, using a method that 
is close to the classical reference probability method of Zakai \cite{Zak69}.

Let us briefly recall the classical procedure of Zakai.  In order to
simplify the filtering problem, one starts by introducing a new measure,
using a Girsanov transformation, under which the observation is a Wiener
process.  Then various (elementary) properties of the conditional
expectation allow the filtering problem to be expressed, and solved, with
respect to the new measure.   Below we will apply this logic to the 
quantum filtering problem.

The following filtering problem is considered in this 
section.  Let $Z_t=A_t+A_t^*$ and denote by $\mathcal{A}_t$ the commutant of 
$\mathcal{C}_t=\mathrm{vN}(Z_s:0\le s\le t)$.  With respect to the
state $\mathbb{P}=\rho\otimes\phi$ on the algebra 
$\mathcal{B}\otimes\mathcal{W}$, we consider the system-observation pair
$$
	dj_t(X)=j_t(\mathcal{L}_{L,H}(X))\,dt+
		j_t([L^*,X])\,dA_t+j_t([X,L])\,dA_t^*
$$
$$
	dY_t=j_t(L+L^*)\,dt+dA_t+dA_t^*
$$
where $j_t(X)=U_t^*XU_t$ and $U_t$ is given by (\ref{eq HP}). 
We are interested in finding a recursive equation for 
$\pi_t(X)=\mathbb{P}(j_t(X)|\mathcal{Y}_t)$ where
$\mathcal{Y}_t=\mathrm{vN}(Y_s:0\le s\le t)$. 

For future reference we first state two elementary properties of 
the conditional expectation.  Let $K_t$ be an adapted process in
$\mathcal{A}_t$.  First, we claim that
$\mathbb{P}(K_s|\mathcal{C}_t)=\mathbb{P}(K_s|\mathcal{C}_s)$.
This property follows from the fact that 
$\mathcal{C}_t=\mathcal{C}_s\otimes\mathcal{C}_{[s,t]}$ and that $K_s$ is 
independent from $\mathcal{C}_{[s,t]}$ by adaptedness.  Second, conditional 
expectations and integrals can be exchanged as follows:
\begin{equation}\label{eq:integrals}
                        \mathbb{P}\left(\left.
                                \int_0^tK_s\,ds\right|
                                \mathcal{C}_t\right)=
                        \int_0^t\mathbb{P}(K_s|\mathcal{C}_s)\,ds,
		\qquad
                        \mathbb{P}\left(\left.
                                \int_0^tK_s\,dZ_s\right|
                                \mathcal{C}_t\right)=
                        \int_0^t\mathbb{P}(K_s|\mathcal{C}_s)\,dZ_s.
\end{equation}
These properties are immediate if $K_t$ is a simple process, and a proof 
of the general case is not difficult.  Note, however, that these properties 
would not be as straightforward if we were to replace $\mathcal{C}_t$ 
by $\mathcal{Y}_t$.  Unlike in the classical It\^o
theory, quantum It\^o theory is grounded in the explicit representation 
of noise processes on Fock space; a concept such as adaptedness is 
defined with respect to the Fock space, rather than with respect to the 
integrator process.  

Though representation-free quantum It\^o theories have been considered in
the literature \cite{AFQ92}, we choose for simplicity to stick to
the Hudson-Parthasarathy theory.  In this context, the discussion in the
previous paragraph suggests that conditioning is most easily performed if
we rotate the problem in such a way that the observations lie entirely in
the Fock space. To this end we use another elementary property of
conditional expectations: if $U$ is a unitary operator and we define a new
state $\mathbb{Q}(X)=\mathbb{P}(U^*XU)$, then
$\mathbb{P}(U^*YU|U^*\mathcal{C}U)= U^*\mathbb{Q}(Y|\mathcal{C})U$ (this
can be verified using the definition of the conditional expectation.) In
our setup, this implies that
$\pi_t(X)=U_t^*\mathbb{Q}^t(X|\mathcal{C}_t)U_t$ where we have defined the
state $\mathbb{Q}^t(X)=\mathbb{P}(U_t^*XU_t)$. 

{\bf Remark.} The time-dependent state $\mathbb{Q}^t$ is precisely the
{\it Schr{\"o}dinger picture} state; in this picture (used in
\cite{BGM04}) the state evolves in time rather than the observables. 
This approach is ubiquitous in quantum mechanics but is less common in
classical stochastics.  Here we prefer to work in the Heisenberg picture;
we only consider the state $\mathbb{Q}^t$ for fixed time $t$ as an
intermediate step in obtaining the Kallianpur-Striebel formula below,
which avoids the technical hurdles due to the explicit representation of
quantum noises on the Fock space. 

We are now ready to get down to business.  Inspired by the classical 
reference probability method, we seek a change of measure (change of 
state) that reduces the filtering problem to elementary manipulations.
The following lemma describes how conditional expectations are related 
under a {\it nondemolition} change of state.

\begin{lem}\label{thm KS}
Let $(\NC, \mathbb{P})$ be a noncommutative probability space
with normal state $\mathbb{P}$.  Let $\CC$ be a commutative von Neumann 
subalgebra of $\NC$ and let $\AC$ be its commutant, i.e.\ 
$\AC = \CC':= \{N\in \NC;\ NC=CN, \  \forall C \in \CC\}$. 
Furthermore, let $V$ be an element in $\AC$ such that $\BB{P}[V^*V]=1$. 
We can define a state on $\AC$ by $\BB{Q}[A] := \BB{P}[V^*AV]$ and we have
  \begin{equation*}
  \BB{Q}\big[X|\CC\big] = \frac{\BB{P}\big[V^*XV|\CC\big]}
  {\BB{P}\big[V^*V|\CC\big]},\ \ \ \ X\in \AC.
  \end{equation*}
\end{lem}
\begin{proof}
Let $K$ be an element of $\CC$. For all $X \in \AC$, 
we can write
  \begin{equation*}\begin{split}
  &\BB{P}\Big[\BB{P}\big[V^*XV|\CC\big]K\Big] = \BB{P}\big[V^*XKV\big] = 
  \BB{Q}\big[XK\big] = \BB{Q}\Big[\BB{Q}\big[X|\CC\big]K\Big] =
  \BB{P}\Big[V^*V\BB{Q}\big[X|\CC\big]K\Big] = \\
  &\BB{P}\Big[\BB{P}\Big[V^*V\BB{Q}\big[X|\CC\big]K\Big|\CC\Big]\Big] =
  \BB{P}\Big[\BB{P}\big[V^*V\big|\CC\big]\BB{Q}\big[X|\CC\big]K\Big],
  \end{split}\end{equation*}
proving the lemma.  
\end{proof}

\textbf{Remark.}  It is essential that $V$ is an element of the commutant of 
$\CC$. The proof would not have worked if $V$ were in $\NC$.  The 
conditional expectation can only be defined from the commutant onto $\CC$ 
and in the first and fourth step of the proof we explicitly used that $V$ 
is an element of $\AC$.  

Our next goal is to find a suitable operator $V$ in order to apply Lemma
\ref{thm KS} to $\mathbb{Q}^t(X|\mathcal{C}_t)$.  Conveniently we have
already expressed $\mathbb{Q}^t(X)$ as $\mathbb{P}(U_t^*XU_t)$, and under
$\mathbb{P}$ we can directly apply the result Eq.\ (\ref{eq:integrals}).
However, $U_t$ is in general not in $\AC_t$, so that Lemma \ref{thm KS}
can not be applied with $V=U_t$.  This is best seen in equation \eqref{eq
HP again}, where aside from $Z_t$ the incompatible noise $i(A_t-A_t^*)$
appears as one of the driving terms in the equation for $U_t$.  The
problem is resolved by the following technique, which to our knowledge
first appeared in a paper by Holevo \cite{Hol90}. 

\begin{lem}
\label{lem:vtdilation}
	Let $V_t$ be the solution of the QSDE
	\begin{equation}
	\label{eq Vt}
		dV_t=\Big\{
			L(dA_t^*+dA_t)-\frac{1}{2}L^*Ldt-iHdt
		\Big\}V_t.
	\end{equation}
	Then $V_t\in\mathcal{A}_t$ and
	$\mathbb{Q}^t(X)=\mathbb{P}(V_t^*XV_t)$.
\end{lem}

\begin{proof}
Suppose without loss of generality that 
$\mathbb{P}(X)=\langle\psi\ten\Phi,X\,\psi\ten\Phi\rangle$
for some vector $\psi\in\mathcal{H}$ (and $\Phi$ is the vacuum vector); as 
$\mathcal{H}$ is finite-dimensional, we can always write any state 
$\rho\otimes\phi$ as a convex combination of vector states of this form.
Now note that
  \begin{equation*}\begin{split}
  &d(U_t\psi\ten\Phi)  = \{LdA_t^* -L^*dA_t - \frac{1}{2}L^*Ldt - iHdt\Big\}U_t\psi\ten\Phi = \\ 
  &(LU_t\psi\ten\Phi_{t]})\ten (dA^*_t\Phi_{[t}) - (L^*U_t\psi\ten\Phi_{t]})\ten (dA_t\Phi_{[t})
  - \frac{1}{2}L^*LU_t\psi\ten\Phi dt - iHU_t\psi\ten\Phi dt = \\
  &\big(LU_t\psi\ten\Phi_{t]}\big)\ten \big(dA^*_t\Phi_{[t}\big) + 
  \big(LU_t\psi\ten\Phi_{t]}\big)\ten \big(dA_t\Phi_{[t}\big)
  - \frac{1}{2}L^*LU_t\psi\ten\Phi dt - iHU_t\psi\ten\Phi dt,
  \end{split}\end{equation*} 
since $dA_t\Phi_{[t}$ is $0$.  It follows that $U_t\,\psi\ten\Phi = 
V_t\,\psi\ten\Phi$, and hence 
$\mathbb{P}(U_t^*XU_t)=\mathbb{P}(V_t^*XV_t)$ for any 
$X\in\mathcal{B}\otimes\mathcal{W}$.  Moreover, $V_t$ is an element of 
$\AC_t$; indeed, equation \eqref{eq Vt} is driven only by a single 
commutative noise $A_t + A_t^*$.  
\end{proof}

Putting these results together, we obtain a noncommutative version of the 
classical Kallianpur-Striebel formula.

{\bf Corollary. (Noncommutative Kallianpur-Striebel)}
	Let $\sigma_t(X)=U_t^*\mathbb{P}(V_t^*XV_t|\mathcal{C}_t)U_t$.	
	Then
	\begin{equation}
		\pi_t(X)=\frac{\sigma_t(X)}{\sigma_t(I)},\qquad
		\forall\,X\in\mathcal{B}.
	\end{equation}

We now obtain an explicit expression for $\sigma_t(X)$.  Using the quantum 
It\^o rules
	$$
  V^*_tXV_t = X + \int_0^tV_s^*\mathcal{L}_{L,H}(X)V_sds + \int_0^tV^*_s(L^*X+XL)V_sd(A_s+A_s^*).
	$$
Using Eq.\ (\ref{eq:integrals}) we obtain
	$$
  \BB{P}\big[V_t^*XV_t|\CC_t\big] = 
  \BB{P}\big[X\big] + 
  \int_0^t\BB{P}\big[V_s^*\mathcal{L}_{L,H}(X)V_s|\CC_s\big]ds + 
  \int_0^t\BB{P}\big[V^*_s(L^*X+XL)V_s|\CC_s\big]d(A_s+A_s^*).
	$$
Another application of the quantum It\^o rules now yields immediately the 
{\it Belavkin-Zakai equation}
\begin{equation}\label{eq Zakai}
	  d\sigma_t(X) = 
	  \sigma_t\big(\mathcal{L}_{L,H}(X)\big)dt + 
	  \sigma_t(L^*X+XL)dY_t.
\end{equation}

{\bf Remark.} It follows from the linearity and complete positivity of 
the conditional expectation that $\sigma_t$ is a linear positive 
functional on $\BC$, i.e.\ apart from the normalisation it is 
a (random) state.  Hence Eq.\ (\ref{eq Zakai}) is a noncommutative 
analogue of the Zakai equation in classical filtering theory and 
propagates the unnormalized conditional state $\sigma_t$.

By applying the (quantum) It\^o rules to the Kallianpur-Striebel formula, 
we obtain an expression for the normalized conditional state
  \begin{equation}\label{eq qfilter}
  d\pi_t(X) = \pi_t\big(\mathcal{L}_{L,H}(X)\big)dt + 
  \Big(\pi_t(L^*X+XL)-\pi_t(L^*+L)\pi_t(X)\Big)
  \Big(dY_t- \pi_t(L^*+L)dt\Big). 
  \end{equation}
This {\it Belavkin-Kushner-Stratonovich quantum filtering equation} is a 
quantum analogue of the classical Kushner-Stratonovich equation.

We conclude this section with an investigation of the
{\it innovations process} $d\overline{Z}_t=dY_t- \pi_t(L^*+L)dt$.
The general theorem below \cite{Bel92b,BGM04} shows that $\overline{Z}_t$ 
is a martingale; but as $d\overline{Z}_t^2=dt$ it must be a Wiener process 
by L\'evy's theorem (the latter can be applied directly if we use the 
spectral theorem to find a classical representation of $\overline{Z}_t$.)
Also this property is in complete analogy with the innovations process in 
classical filtering theory.

\begin{stel}
\label{thm:innovation}
Let $dZ_t = a_td\Lambda_t + b_tdA^*_t + b^*_tdA_t$ and $Z_0 =0$ 
where $a_t \in \BB{R}$ and $b_t \in \BB{C}$ for all $t\ge 0$.
Define the \emph{innovating martingale} $\overline{Z}_t$ by 
  \begin{equation*}
  \overline{Z}_t := U_t^* Z_tU_t - \Big(\int_0^t a_s\pi_s(L^*L) +
  b_s\pi_s(L^*)+ b^*_s\pi_s(L)ds\Big).
  \end{equation*}
Then $\overline{Z}_t$ is a martingale, i.e.\ for all $t\ge s\ge 0$ we have 
$\BB{P}[\overline{Z}_t|\mathcal{Y}_s] = \overline{Z}_s$.
\end{stel}

\begin{proof}
We need to prove that 
$\BB{P}[\overline{Z}_t-\overline{Z}_s|\mathcal{Y}_s] = 0$
for all $t \ge s \ge 0$.  This means we 
have to prove for all $t \ge s \ge 0$ and $K \in \mathcal{Y}_s$
that $\BB{P}\big[(\overline{Z}_t-\overline{Z}_s)K\big] = 0$. This is 
equivalent to 
  \begin{equation*}
  \mathbb{P}(U_t^*Z_tU_tK) - \mathbb{P}(U^*_sZ_sU_sK) = 
  \int_s^t \mathbb{P}\Big(a_r\pi_r(L^*L)K +
	b_r\pi_r(L^*)K+ b^*_r\pi_r(L)K\Big)ds,
  \end{equation*}
for all $t\ge s\ge 0$ and $K\in \mathcal{Y}_s$. 
Since $K \in \mathcal{Y}_s$ it can be written as 
$K =U_s^*CU_s = U_t^*CU_t$ for some $C \in \CC_s$. 
Furthermore, if $C$ runs through $\CC_s$, $K$ runs 
through $\mathcal{Y}_s$. For $t=s$ the above 
equation is true, therefore it remains to show that 
for all $C \in \CC_s$
  \begin{equation*}\begin{split}
  d\mathbb{P}\big(U_t^*Z_tCU_t\big) & = 
\mathbb{P}\big(a_r\pi_r(L^*LC) +
b_r\pi_r(L^*C)+ b^*_r\pi_r(LC)\big)dt  \\
  & = \mathbb{P}\big(U_t^*(a_rL^*LC + b_rL^*C + 
\overline{b}_rLC)U_t\big)dt.
  \end{split}\end{equation*}
But this is just an exercise in applying the quantum It\^o rules.
\end{proof}

\section{More examples}\label{sec examples}

\subsection{Controlled quantum diffusions}

Up to this point we have considered filtering only for simple quantum
systems interacting with a field.  However, an important application of
quantum filtering theory is in {\it quantum control}---we can feed back a
signal to the initial system based on the observations we have made, in
real time, in order to achieve some particular control goal.  In control
theory, the filter is often used as an intermediate step in generating the
control signal.  Before we can do this, however, we must show that the
filtering equations retain their form even in the presence of feedback.  
This can be done in a natural way in the reference probability approach.

The basic object we need is a {\it controlled quantum diffusion}.  As 
before, we fix the algebra $\BC\otimes\WC$.  The controlled diffusion 
consists of two things:
\begin{itemize}
\item The output noise $Z_t$ (we will choose $Z_t=A_t+A_t^*$).
\item The controlled Hudson-Parthasarathy QSDE\footnote{
        For existence and uniqueness, see e.g.\ \cite{Lindsay}.
}
        $$
          dU_t=\left\{
                L_t\,dA_t^*-L_t^*\,dA_t-\frac{1}{2}L_t^*L_t\,dt
                        -iH_t\,dt
          \right\}U_t
        $$
        where for all $t$, $H_t$ and $L_t$ are affiliated to
        $\BC\otimes\CC_t$ ($\CC_t={\rm vN}(Z_s:0\le s\le t)$),
        and $H_t$ is self-adjoint.      
\end{itemize}
As before, we find that this model gives rise to the system-observation 
pair
$$
        dj_t(X)=j_t(\mathcal{L}_{L_t,H_t}(X))\,dt+
        j_t([L_t^*,X])\,dA_t+j_t([X,L_t])\,dA_t^*
$$
$$
        dY_t=j_t(L_t+L_t^*)\,dt+dA_t+dA_t^*
$$
where $j_t(X)=U_t^*XU_t$ is the flow and $Y_t=U_t^*Z_tU_t$ is the output 
process.  

To expose the connection with control, consider the simplest case where
$L_t=L\in\BC$ is fixed and $H_t=H\otimes u_t(Z_{s\le t})$, i.e.\ $H$ is a
fixed Hamiltonian in $\BC$ and $u_t$ is some real function of the history 
of the output noise up to time $t$.  Then we get
$$
        dj_t(X)=u_t(Y_{s\le t})\,j_t(i[H,X])\,dt
                +j_t(\mathcal{L}_{L,I}(X))\,dt+
                j_t([L^*,X])\,dA_t+j_t([X,L])\,dA_t^*
$$
$$
        dY_t=j_t(L+L^*)\,dt+dA_t+dA_t^*
$$
where it is important to note that by pulling it outside $j_t(\cdot)$,
$u_t$ becomes a function of the ouput process! This is just the model we
considered before, with the difference that here the Hamiltonian term is 
modulated by a control signal which is some (arbitrary) function of the
observations history.  This is precisely the situation in (Hamiltonian)
feedback control, the idea being that we choose the function $u_t$ so that
a particular control goal is achieved.  The general model of a controlled
quantum diffusion allows us to make both $H$ and $L$ arbitrary functions
of the observations history, and thus provides a general model for quantum
systems with feedback.

The question we ask here is, what is the form of the filter when we allow 
for feedback?  It turns out that nothing much changes in the derivation of 
the filter using the reference probability method.  We can follow all the 
same steps to obtain the Kallianpur-Striebel formula
$$
        \pi_t(X)=\mathbb{P}(j_t(X)|{\rm vN}(Y_s:0\le s\le t))=
        \frac{\sigma_t(X)}{\sigma_t(I)},\qquad
        \sigma_t(X)=U_t^*\mathbb{P}(V_t^*XV_t|\CC_t)U_t
$$
with the $\CC_t'$-affiliated change of measure
$$
        dV_t=\left\{L_t\,dZ_t-\frac{1}{2}L_t^*L_t\,dt
                -iH_t\,dt
        \right\}V_t.
$$
By applying the quantum It\^o rules, we now obtain the Zakai equation
$$
        d\sigma_t(X)=\sigma_t(\LC_{L_t,H_t}(X))\,dt+
                \sigma_t(L_t^*X+XL_t)\,dY_t.
$$
In particular, suppose that as before $L_t=L$, $H_t=H\otimes u_t(Z_{s\le 
t})$.  Note that $\mathbb{P}(V_t^*i[H_t,X]V_t|\CC_t)=
u_t(Z_{s\le t})\,\mathbb{P}(V_t^*i[H,X]V_t|\CC_t)$ as $V_t$ commutes with 
$Z_{s\le t}$ and by the module property of the conditional expectation.  
Thus in this case
$$
        d\sigma_t(X)=
                u_t(Y_{s\le t})\,\sigma_t(i[H,X])\,dt+
                \sigma_t(\LC_{L,I}(X))\,dt+
                \sigma_t(L^*X+XL)\,dY_t.
$$
We see that as expected, the filter for a quantum diffusion with feedback 
has precisely the same form as the filter obtained previously with 
feedback added in a naive way.  This provides a foundation for the 
use of the filter in quantum feedback control.

\subsection{Imperfect observations}

We consider the same filtering problem as in the section 
\ref{sec KS formula}, except that now we add additional corrupting noise 
to the observation
$$
	dY_t=j_t(L+L^*)\,dt+dA_t+dA_t^*+\kappa(dB_t+dB_t^*).
$$
Here $\kappa\ge 0$ and $B_t$ is an additional field, the {\it corrupting 
noise}, which is independent of $A_t$; i.e., we have tensored another 
copy of the Fock space $\mathcal{F}$ onto the Hilbert space, and $B_t$ is 
the fundamental process on this space.  We now take 
$Z_t=A_t+A_t^*+\kappa(B_t+B_t^*)$; as before
$\mathcal{C}_t=\mathrm{vN}(Z_s:0\le s\le t)$ and $\mathcal{A}_t$ is its 
commutant.  Note that we still have $Y_t=U_t^*Z_tU_t$, etc.

Up to and including the Kallianpur-Striebel formula, this case is 
identical to the one treated in section \ref{sec KS formula}.  As 
before, we have
	$$
  V^*_tXV_t = X + \int_0^tV_s^*\mathcal{L}_{L,H}(X)V_sds + \int_0^tV^*_s(L^*X+XL)V_sd(A_s+A_s^*).
	$$
However, we cannot directly apply Eq.\ (\ref{eq:integrals}) at this 
point, as in this case the stochastic integral with respect to $A_s+A_s^*$ 
does not correspond to the integral with respect to $Z_s$.  Let us write
	$$
		A_t+A_t^*=\alpha Z_t
		+\left[(1-\alpha)(A_t+A_t^*)-\alpha\kappa(B_t+B_t^*)\right],
	$$
which holds for any $\alpha\in\mathbb{R}$.  Now note that if we choose 
the particular value $\alpha=(1+\kappa^2)^{-1}$, then the noise $Z_t$ is 
independent from $M_t=(1-\alpha)(A_t+A_t^*)-\alpha\kappa(B_t+B_t^*)$.
By an elementary argument, it follows that
	$$
                        \mathbb{P}\left(\left.
                                \int_0^tK_s\,dM_s\right|
                                \mathcal{C}_t\right)=0.
	$$
This, together with Eq.\ (\ref{eq:integrals}), directly gives the 
Belavkin-Zakai equation with imperfect observations
\begin{equation}
	  d\sigma_t(X) = 
	  \sigma_t\big(\mathcal{L}_{L,H}(X)\big)dt + 
	  (1+\kappa^2)^{-1}\sigma_t(L^*X+XL)dY_t.
\end{equation}

\subsection{Photon counting observations}

Once again we consider the same system; instead of homodyne detection, 
however, we now perform photon counting in the field: i.e.,
$Y_t=U_t^*\Lambda_tU_t$ which gives
$$
	dY_t=d\Lambda_t+j_t(L)dA_t^*+j_t(L^*)dA_t+j_t(L^*L)dt.
$$
We would like to follow the same procedure as for homodyne detection.  The 
following lemma, which replaces Lemma \ref{lem:vtdilation}, suggests how 
to proceed.  The proof is identical to that of Lemma \ref{lem:vtdilation}.

\begin{lem}
\label{lem:countvt}
	Let $U_t'$ be the solution of the Hudson-Parthasarathy equation
	$$
		dU_t'=\Big\{L'dA^*_t-L^{\prime *}dA_t-\frac{1}{2}
			L^{\prime *}L'dt-iH'dt
		\Big\}U_t'
	$$
	and let $V_t'$ be the solution of
	$$
		dV_t'=\Big\{L'(d\Lambda_t+dA^*_t+dA_t+dt)
			-\frac{1}{2}L^{\prime *}L'dt-L'dt-iH'dt
		\Big\}V_t'.
	$$
	Then $V_t\in\mathrm{vN}(\Lambda_s+A^*_s+A_s+s:0\le s\le t)$
	and $\mathbb{P}({U_t'}^*XU_t')=\mathbb{P}({V_t'}^*XV_t')$.
\end{lem}

Define $Z_t=\Lambda_t+A_t^*+A_t+t$, and as before $\mathcal{A}_t$ is the
commutant of $\mathcal{C}_t=\mathrm{vN}(Z_s:0\le s\le t)$ and
$\mathcal{Y}_t=\mathrm{vN}(Y_s:0\le s\le t)$.  Lemma \ref{lem:countvt} 
directly provides us with a nondemolition change of measure, provided 
that we rotate our problem so that $\mathcal{Y}_t={U_t'}^*\mathcal{C}_tU_t'$
using a suitable unitary $U_t'$.  Then, defining $\sigma_t(X)=
{U_t'}^*\mathbb{P}({V_t'}^*XV_t'|\mathcal{C}_t)U_t'$, the 
Kallianpur-Striebel formula holds for $\sigma_t(X)$.

Define $W_t$ as the solution of the QSDE
$$
	dW_t=[dA_t-dA_t^*-\tfrac{1}{2}dt]W_t
$$
(compare with Eq.\ (\ref{eq dif Weyl}).)  Using the quantum It\^o rules 
one can verify that $\Lambda_t=W_t^*Z_tW_t$.  But recall that 
$Y_t=U_t^*\Lambda_tU_t=U_t^*W_t^*Z_tW_tU_t$; thus $U_t'=W_tU_t$ is our 
rotation of choice.  Another application of the quantum It\^o rules gives
$$
	dU_t'=\Big\{
		(L-1)dA_t^*-(L^*-1)dA_t-\frac{1}{2}
		(L^*L+I-2L+2iH)dt
	\Big\}U_t',
$$
which corresponds to the nondemolition change of measure
$$
	dV_t'=\Big\{
		(L-1)dZ_t-\frac{1}{2}(L^*L-I+2iH)dt
	\Big\}V_t'.
$$
For $X\in\mathcal{B}$, we obtain using the quantum It\^o rules
  \begin{equation*}
  d{V_t}'^*XV_t' = {V_t'}^* \big(\mathcal{L}_{L,H}(X)\big)V_t'dt + 
	{V_t'}^*(L^*XL-X)V_t'(dZ_t-dt). 
  \end{equation*}  
Finally we obtain using the definition of $\sigma_t$, Eq.\ 
(\ref{eq:integrals}), and the quantum It\^o rules
  \begin{equation}\label{eq linBel}
  d\sigma_t(X) = \sigma_t\big(\mathcal{L}_{L,H}(X)\big)dt + 
  \big(\sigma_t(L^*XL)-\sigma_t(X)\big)\big(dY_t-dt\big). 
  \end{equation}
which is the Belavkin-Zakai equation for counting observations.

Using the Kallianpur-Striebel formula we can now obtain an expression for 
the normalized conditional state
  \begin{equation*}
  d\pi_t(X) = \pi_t\big(\mathcal{L}_{L,H}(X)\big)dt + 
  \Big(\frac{\pi_t(L^*XL)}{\pi_t(L^*L)}-\pi_t(X)\Big)
	\big(dY_t-\pi_t(L^*L)dt\big), 
  \end{equation*}
which is the quantum filtering equation for photon counting.  By Theorem 
\ref{thm:innovation}, we see that the innovations process
$d\overline{Z}_t=dY_t-\pi_t(L^*L)dt$ is a martingale.  Note that this 
implies that in terms of the conditional state, $\overline{Z}_t$ is a 
counting process with rate $\pi_t(L^*L)$.

\subsection{Other filters}

There is a wide array of other relevant examples, which we do not 
discuss here.  For risk-sensitive filtering equations and their use in 
risk-sensitive control we refer to \cite{Jam04,Jam05}.  Moreover, sometimes
the input noise is not taken to be in the vacuum state, as is the case for
squeezed or thermal noise.  For a general treatment we refer to 
\cite{Bel92a}, and for the specific case of squeezed noise see \cite{Bou04b}.

\section*{Acknowledgment}

This work was supported by the ARO under Grant DAAD19-03-1-0073.

\bibliography{ref}
\end{document}